\documentclass[fleqn,usenatbib]{mnras}

\usepackage{newtxtext,newtxmath}
\usepackage{pdflscape}	

\usepackage[T1]{fontenc}

\DeclareRobustCommand{\VAN}[3]{#2}
\let\VANthebibliography\thebibliography
\def\thebibliography{\DeclareRobustCommand{\VAN}[3]{##3}\VANthebibliography}

\newcommand{\hiMpc}{h^{-1}~{\rm Mpc}}
\newcommand{\hikpc}{h^{-1}~{\rm kpc}}
\newcommand{\vpeak}{V_{\rm peak}}
\newcommand{\zprog}{z_{\rm prog}}
\newcommand{\Mprog}{M_{\rm prog}}
\newcommand{\hiMsun}{h^{-1}~M_\odot}
\newcommand{\ngal}{n_{\rm gal}}

\newcommand{\rd}{{\rm d}}
\newcommand{\fsat}{f_{\rm sat}}
\newcommand{\wpp}{w_{\rm p}}

\usepackage{color}

\newcommand{\smrv}[1]{\textcolor{black}{#1}}
\newcommand{\smrvv}[1]{\textcolor{black}{#1}}

\usepackage{ulem}

\usepackage{graphicx}	
\usepackage{amsmath}	

\title[\smrv{Progenitor matching finds two stellar mass growth modes}]
{Subhalo abundance matching using progenitor mass \smrv{at varying redshift: Two modes of stellar mass growth imprinted into the Subaru HSC galaxy clustering}}

\author[S.~Masaki, D.~Kashino, \smrv{S.~Ishikawa} and Y.-T.~Lin]
{Shogo~Masaki$^{1,2}$ \thanks{shogo.masaki@gmail.com}, Daichi~Kashino$^{3,2}$, \smrv{Shogo~Ishikawa$^{4}$} and  
Yen-Ting Lin$^{5}$\\
$^{1}$National Institute of Technology, Suzuka College, Suzuka, Mie 510-0294, Japan\\
$^{2}$Department of Physics, Nagoya University, Nagoya, Aichi 464-8601, Japan\\
$^{3}$Institute for Advanced Research, Nagoya University, Nagoya, Aichi 464-8601, Japan\\
\smrv{$^4$Center for Gravitational Physics and Quantum Information, Yukawa Institute for Theoretical Physics, Kyoto University, Sakyo-ku, Kyoto 606-8502, Japan}\\
$^{5}$Institute of Astronomy and Astrophysics, Academia Sinica  (ASIAA),  Taipei 10617, Taiwan}

\date{Accepted XXX. Received YYY; in original form ZZZ}

\pubyear{2023}

\begin{document}
\label{firstpage}
\pagerange{\pageref{firstpage}--\pageref{lastpage}}
\maketitle

\begin{abstract}
We propose a novel subhalo abundance matching (SHAM) model that uses the virial mass of the main progenitor of each (sub)halo $M_{\rm prog}$ as a proxy of the galaxy stellar mass $M_*$ at the time of observation.  
This $M_{\rm prog}$ model predicts the two-point correlation functions depending on the choice of the epoch $z_{\rm prog}$ at which $M_\mathrm{prog}$ is \smrv{quoted}.
With $z_{\rm prog}$ as a fitting parameter, we apply the $M_{\rm prog}$ model to the angular correlation functions \smrv{measured with varying stellar mass thresholds from $M_{*,~{\rm lim}}/(h^{-2}M_\odot)=10^{11}$ to $10^{8.6}$ using a sample of galaxies at $z\simeq0.4$ from the Subaru Hyper Suprime-Cam survey}.
The $M_{\rm prog}$ model can reproduce the observations very well \smrv{over $10~\hikpc\textrm{--}10~\hiMpc$}.
We find that, for the samples of $10^{9.2}\leq M_{*,~{\rm lim}}/(h^{-2}M_\odot)\leq10^{10.2}$, the correlation functions predicted by the widely-used $V_{\rm peak}$ model lack amplitudes at $\lesssim1~\hiMpc$, \smrv{suggesting that $M_{\rm prog}$ is a better proxy of the galaxy stellar mass than conventional $V_{\rm peak}$}.
The $z_{\rm prog}$ parameter is highest (\smrv{$z_{\rm prog}\simeq3$}) for intermediate mass galaxies at \smrv{$M_*\simeq10^{9.9}~h^{-2}M_\odot$}, and becomes smaller \smrv{down to} $z_\mathrm{prog}\simeq1$ for both lower- and higher-mass galaxies.
We interpret these trends as reflecting the downsizing in the in-situ star formation in lower-mass galaxies and the larger contribution of \smrv{the} ex-situ stellar mass growth in higher-mass galaxies.

\end{abstract}




\section{Introduction}
The observed spatial distribution of galaxies contains a huge amount of physical information on cosmology, and galaxy formation and evolution.
To extract such information accurately from modern wide-field galaxy surveys, it is necessary to correctly model the relations between the observed galaxies and their host dark matter structures, i.e., halos and their substructures, subhalos\footnote{For simplicity, unless explicitly noted, hereafter we shall refer to all halos and subhalos simply as subhalos.}.

One of the empirical methods widely used for modeling the galaxy-subhalo connection is the subhalo abundance matching (SHAM) method (\citealt{Kravtsov04}; see \citealt{WechslerTinker18} for a recent review).
The SHAM method assumes a monotonic relation with some scatter between a galaxy observable and a simulated subhalo property.
In the simplest form, known as the rank-ordering SHAM, for a sample of observed galaxies selected by a threshold in some physical property (e.g., stellar mass or luminosity), the corresponding subhalo sample is constructed by taking the threshold of the selected property so that the subhalo number density matches that of the galaxy sample.
Unlike the halo occupation distribution (HOD) method \citep[see][for a review]{Cooray02}, which is also a frequently used method, SHAM can incorporate subhalo distributions on small scales realized in cosmological $N$-body simulations.

In performing SHAM, one should use a subhalo property that is expected to strongly correlate with the target galaxy property.
The choice of such a property would have a profound impact on the predicted galaxy statistics, including clustering measurements \citep[e.g.,][]{conroy06,Behroozi13_wc,Reddick13,Moster13,Chaves-Montero16,Lehmann17,Moster18,Behroozi19,Stiskalek21,Tonnesen21,Chuang23,Contreras23}.
Thus, comparing the SHAM predictions with observational results can validate the chosen subhalo property.
\cite{Reddick13} used various subhalo properties to study how the predictions for galaxy statistics at $z\simeq0$ are affected.
They showed that using the peak maximum circular velocity of each subhalo in its lifetime, $V_{\rm peak}$, can reproduce the observed projected correlation functions (PCFs) of the galaxy samples constructed with a threshold of stellar mass and luminosity.
Hereafter, we refer to this rank-ordering approach as the $\vpeak$ model.

However, the validity of using $\vpeak$ as a proxy for stellar mass or luminosity is still unclear, despite its successful reproduction of observed two-point correlation functions (2PCFs) of some types of galaxies \citep{Nuza13,RT16,Saito16,Alam17,DongPaez22}.
Since satellite subhalos reach $\vpeak$ before getting accreted onto host halos as shown by \cite{Behroozi14}, the physical relation between $\vpeak$ and star formation activity in satellite galaxies is uncertain.
\cite{Campbell18} argued that stellar mass growth implied by the $\vpeak$ model is too early.
The implied early growth does not agree with the mass-based SHAM models tuned to match the observed evolution of stellar mass functions \citep{Yang12,Behroozi13_wc,Moster13}.
This inconsistency is due to the earlier formation of the gravitational potential well of subhalos than mass \citep{vdB14}.
Interestingly, the rank-ordering model using the peak mass $M_{\rm peak}$ predicts the growth history similar to the stellar mass function-tuned models.
Furthermore, \cite{Leauthaud17} reported that SHAM and HOD models, including the $\vpeak$ model \citep{Reid14,RT16,Saito16,Alam17}, can reproduce the observed PCF of the CMASS galaxies \citep{Ahn14} but overpredict the lensing profile.
\smrv{There are attempts to reconcile the inconsistency between clustering and lensing by incorporating assembly bias \citep{Wechsler06,Gao07} in the HOD modeling, adopting cosmology not derived from the cosmic microwave background measurements, re-estimating the lensing measurements, and combinations of these \citep{Leauthaud17, Lange19, Yuan20, Lange21,Amodeo21,Amon23}.}
However, the situation has not been settled yet.
\cite{Hearin13b} showed that the SHAM models including the $\vpeak$ model exhibit tensions with the observed luminosity functions of field galaxies and group member galaxies. 

\cite{Campbell18} also studied galaxy clustering predicted by the mass-based SHAM models including the rank-ordering model using the peak mass $M_{\rm peak}$, and more detailed models for the stellar mass-halo mass relation \citep{Yang12,Behroozi13_wc,Moster13}.
They showed that these models do not reproduce the observed galaxy clustering without introducing assembly bias as the secondary subhalo property \citep{Masaki13b,Hearin13} or invoking the usage of substantial `orphans', i.e., galaxies hosted by subhalos that fall below the numerical resolution limit.

In this paper, we propose a novel rank-ordering mass-based SHAM model.
Our model uses the virial mass of the progenitor at a redshift $\zprog$ of each subhalo, $\Mprog$, as a proxy of the galaxy stellar mass at the time of  observation.
We refer to this model as the $\Mprog$ model.
\smrv{In this model, $\zprog$ is the only primary parameter and marks the characteristic epoch of stellar mass growth.}
We show that our $M_{\rm prog}$ model with a certain choice of $z_{\rm prog}$ can reproduce the observed 2PCFs with the same or higher amplitudes than the $V_{\rm peak}$ model, without the need to invoke orphan galaxies or any secondary subhalo properties.
We apply the $M_{\rm prog}$ and $V_{\rm peak}$ models to the angular correlation functions (ACFs) of the galaxy samples with several stellar mass thresholds at $z\simeq0.4$ obtained from the Subaru Hyper-Suprime Cam (HSC) survey \citep{Aihara18,Ishikawa20}.
For the samples with the thresholds of $\smrv{9.2}\leq\log_{10}[M_{*,~{\rm lim}}/(h^{-2}M_\odot)]\leq10.2$, we find that the predictions of the $M_{\rm prog}$ model agree better with the observation than the $V_{\rm peak}$ model at $\lesssim1~\hiMpc$.
\smrv{We also find that the dependence of $\zprog$ on stellar mass is qualitatively consistent with the two-phase scenario of stellar mass growth in galaxies, i.e., in-situ star formation and ex-situ star accretion \citep{Oser10}}.

This paper is structured as follows.
In Sec.~\ref{sec:method}, we describe the simulation\smrv{s} used in this paper.
Then we introduce the $\Mprog$ model, study how this model predicts the 2PCFs, \smrv{measure the ACFs of the HSC galaxies,} and discuss how to fit the predictions to the observed ACFs with the stellar mass thresholds.
Sec.~\ref{sec:result} presents the fitting results on  the observed ACFs by both $\Mprog$ and $\vpeak$ models, the \smrv{parameter constraints}, and the inferred satellite fraction.
We summarize our results and conclude in Sec.~\ref{sec:conclusion}.

\section{Methods}
\label{sec:method}
We first present the details of the simulations used in this work,
then describe the SHAM model that uses the virial mass of the progenitor at redshift $\zprog$ of each subhalo, $\Mprog$.
We study the 2PCFs predicted by the $\Mprog$ model and interpret their $\zprog$ dependence.
\smrv{We remeasure the ACFs of the HSC galaxies using the same galaxies and random points data as \cite{Ishikawa20}.}
We conclude this section by discussing how we fit the model predictions to observed clustering.

\subsection{The mini-Uchuu \smrv{and Shin-Uchuu} simulations}
We use the publicly available halo/subhalo catalogs produced from the mini-Uchuu \smrv{and Shin-Uchuu} simulations \citep{Ishiyama21} carried out with the {\sc GreeM} code \citep{ishiyama09}.
The simulations adopt the {\it Planck} 2018 $\Lambda$-cold dark matter ($\Lambda$CDM) cosmological parameters as $\Omega_{\rm m}=0.3089,~\Omega_\Lambda=0.6911,~h=0.6774,~\sigma_8=0.8159,~\Omega_{\rm b}=0.0486$ and $n_{\rm s}=0.9667$, where $h$ is the dimensionless Hubble constant defined by $H_0=100h~{\rm km~s^{-1}~Mpc^{-1}}$ \citep{Planck18}.
Other aspects of the simulations are summarized as follows \smrv{for the mini and Shin runs, respectively}: the number of simulation particles: $N_{\rm part}=2560^3$ \smrv{and $6400^3$}, the simulation box length: $L_{\rm box}=400~\hiMpc$ \smrv{and $140~\hiMpc$}, the softening length: $\epsilon=4.27~\hikpc$ \smrv{and $0.4~\hikpc$}, and the mass of a simulation particle: $m_{\rm part}=3.27\times10^8~\hiMsun$ \smrv{and $8.97\times10^5~\hiMsun$}.
\smrv{The mini-Uchuu simulation offers higher statistical precision, while the Shin-Uchuu simulation enables resolution studies \citep{Guo14, vdB18, Masfield21}.}
The halos and subhalos are identified by the {\sc Rockstar} finder \citep{Behroozi:2013}.
\smrv{To reduce computational costs with the Shin-Uchuu simulation, we use the subhalos with the maximum circular velocity $V_{\rm max}\geq14.27~{\rm km~s^{-1}}$, which is the minimum value in the catalog of the mini-Uchuu simulation.}
We utilize the {\sc nbodykit} package \citep{Hand18} to handle the halo/subhalo catalogs.

\subsection{Our SHAM model}
\label{sec:Mprog}

\subsubsection{Motivation}
\label{sec:motivation}
It is known as the downsizing scenario that higher-mass galaxies tend to assemble their stellar mass and cease star formation at earlier epochs, while lower-mass ones tend to continue star formation to later times \citep{Cowie96, Guzman97, Brinchmann00, Kodama04, Bell05, Jimenez05, Juneau05, Bundy06, Neistein06}.
In this picture, it is naturally expected that the stellar mass of  massive galaxies is better correlated with their progenitors' host subhalo mass at earlier epochs and vice versa.
In other words, we expect that higher-mass galaxies reside in subhalos that were sufficiently massive at higher-$z$.

The above downsizing-based expectation would be the case for galaxies whose stellar mass growth is dominated by the ``in-situ'' star formation.
For most massive galaxies, we also need to account for stellar mass growth via  galaxy mergers, i.e., the ``ex-situ'' star accretion \citep{Oser10, Lackner12, Pillepich15, RG16, Pillepich18, Davison20, Cannarozzo23}.
The process of galaxy mergers increases both the host subhalo and stellar masses.
Cosmological simulations of galaxy formation suggest that the fraction of the ex-situ stars in more massive galaxies increases rapidly and can be dominant over or comparable to those formed in-situ at later epochs \citep{RG16}.
We thus expect that the observed stellar mass of most massive galaxies is represented better by the subhalo mass at epochs nearer the time of observation, instead of some earlier epochs as expected from the downsizing scenarios.

Motivated by these expected correlations between the observed stellar mass and the host subhalo mass, we propose a novel SHAM model.
Our model uses the progenitor subhalo mass $M_\mathrm{prog}$ at an epoch $z=z_\mathrm{prog}$ as a proxy of the observed stellar mass, where $z_\mathrm{prog}$ can vary as a function of the stellar mass.

As we shall see later in Sec.~\ref{sec:xi_zprog}, the predicted 2PCFs depend on the choice of $\zprog$ non-trivially in amplitude and shape.
Treating $\zprog$ as a free parameter in fitting to observed clustering measurements, the obtained best-fit $\zprog$ values should reflect the characteristic epoch of stellar mass growth as a function of the galaxy stellar mass \smrv{at the time of observation}.

As we discussed above, the best-fit $\zprog$ values are expected to be lower toward the higher and lower stellar mass ends and have a peak at the intermediate mass range.
The lower-mass and higher-mass sides of the peak reflect downsizing in-situ star formation and ex-situ star accretion, respectively.

\subsubsection{The implementation}
We now describe the implementation of the $\Mprog$ model.
Among the various definitions of the subhalo mass, we use the virial mass given by the {\sc Rockstar} halo finder as the progenitor mass $\Mprog$.
We fit the model predictions to observed galaxy clustering with two free parameters.
The primary parameter is $\zprog$, the redshift at which we evaluate the virial mass of the progenitor $\Mprog$.

The second parameter is to control the scatter between $M_*$ and $\Mprog$, $\sigma_M$.
Although we assume a tight correlation between the two, there could be a non-negligible scatter in the relation.
To incorporate such a scatter, we perturb $\Mprog$ by multiplying the logarithm of $\Mprog$ with a random number drawn from a Gaussian distribution $\mathcal{N}$ with the mean of $0$ and the standard deviation of $\sigma_M$ \citep{RT16,Yu22} as
\begin{align}
    \log_{10}M_{\rm pert}=[1+\mathcal{N}(0,~\sigma_M)]\log_{10}\Mprog.
    \label{eq:scatter_Mprog}
\end{align}
Such a perturbation leads to more low-mass subhalos in the resultant subhalo samples for larger $\sigma_M$ because they are more abundant than high-mass ones.
Hence, a larger $\sigma_M$ suppresses the overall amplitude of 2PCFs.

We construct the mass accretion histories \citep[MAHs; see e.g.,][]{Wechsler02,McBride09} of the most massive progenitors (MMPs) to evaluate $\Mprog$ using the halo merger trees obtained with the {\sc ConsistentTrees} code \citep{Behroozi13_consistenttree}.
The MAHs of MMPs can be seen as the main trunk of each merger tree.
The available number of the model parameter $\zprog$ is limited by \smrv{the number of output snapshots of the simulations}, i.e., 50 outputs from $z=13.93$ to $z=0$ \smrv{for the mini run, and 70 outputs from $z=19.96$ to $z=0$ for the Shin run}.

The $\Mprog$ model is similar to the SHAM model for luminous red galaxies (LRGs) developed by \cite{masaki13}.
Following observational suggestions on the growth of LRGs, they assumed that the most massive distinct halos at $z=2$ are the progenitors of LRGs, and identified their descendants at $z\simeq0.3$ as LRGs.
They found that their model reproduces observed clustering and lensing profiles qualitatively well.
Our $\Mprog$ model differs in the inclusion of satellite subhalos at $\zprog$ and has higher flexibility as $\zprog$ is a model parameter.

\subsection{The impact of $\zprog$ on 2PCFs}
\label{sec:xi_zprog}
We study the impacts of $\zprog$ on predicting galaxy clustering.
We use the {\sc CorrFunc} package \citep{Sinha19,Sinha20} to measure the real-space 2PCF $\xi$ as a function of the comoving distance $r$.
For this, we construct subhalo samples at $z=0$ \smrv{in the mini-Uchuu simulation} by taking the threshold $\Mprog$ value with several $\zprog$ so that the number density equals to $\ngal=10^{-2}~h^3~{\rm Mpc}^{-3}$.
We estimate the error bars for 2PCFs by the `omit-one' jackknife resampling using the $27$ subvolumes.
We compare the results with those from the $\vpeak$ model, which are taken as the fiducial.
Below we denote the 2PCFs from the $\Mprog$ and $\vpeak$ models as $\xi_M$ and $\xi_V$, respectively.
For simplicity, we do not perturb $\Mprog$ and $\vpeak$.

\begin{figure}
	\includegraphics[width=\columnwidth]{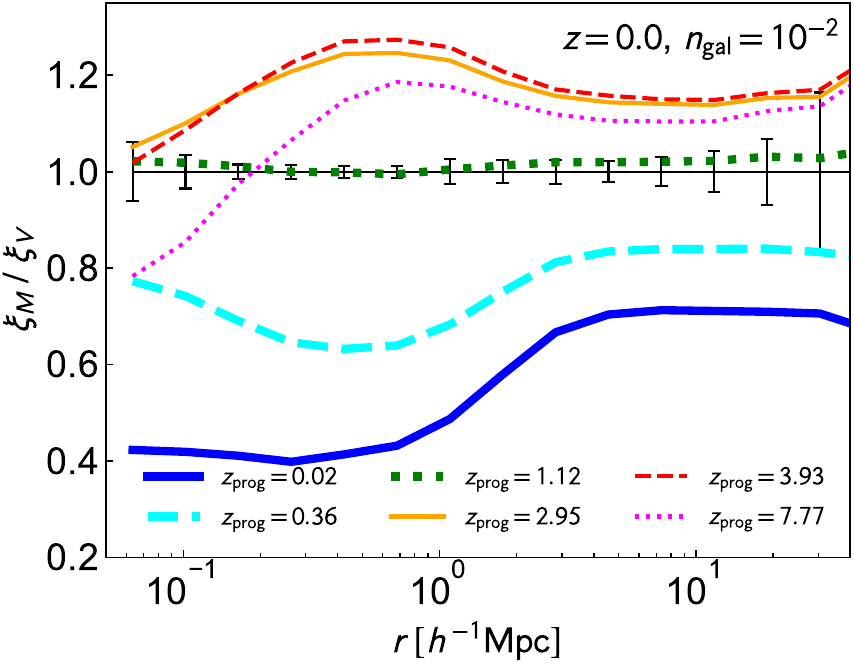}
    \caption{The impact of $\zprog$ on $\xi_M$ for the sample with $n_{\rm gal}=10^{-2}~h^3~{\rm Mpc}^{-3}$ at $z=0$.
    $\xi_M$ is scaled by $\xi_V$.
    The horizontal thin solid line represents $\xi_V$ with the scaled error bars.
    }
    \label{fig:xi_diff_z0p0}
\end{figure}

\begin{figure}
	\includegraphics[width=\columnwidth]{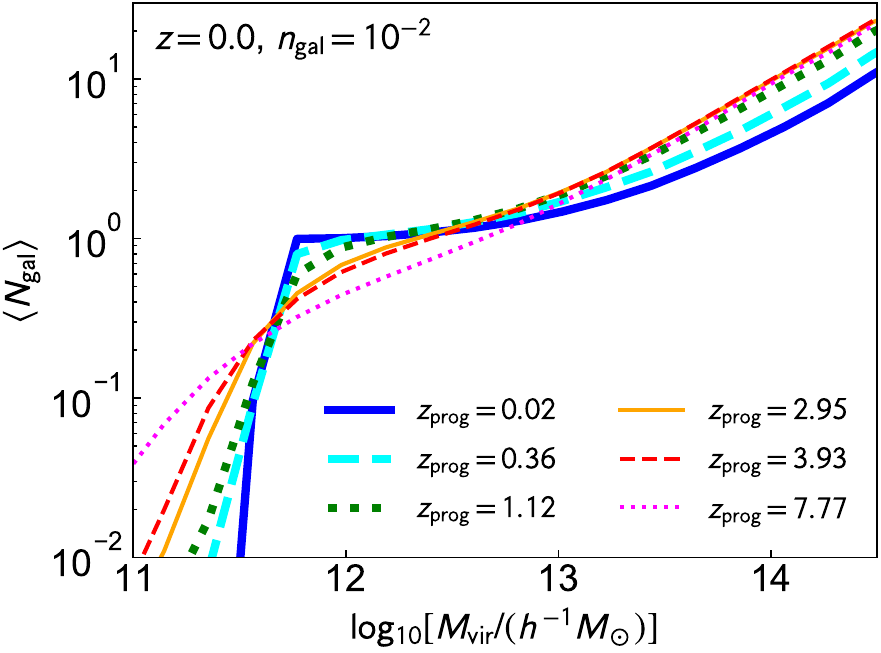}
    \caption{The halo occupation numbers $\langle N_{\rm gal}\rangle$ of the sample with $n_{\rm gal}=10^{-2}~h^3~{\rm Mpc}^{-3}$ at $z=0$ from the $\Mprog$ model with varying $\zprog$.
    }
    \label{fig:hod_Mzprog_z0p0}
\end{figure}
Fig. \ref{fig:xi_diff_z0p0} shows the impact of $\zprog$ on $\xi_M$ at $z=0$.
For ease of comparison, we show $\xi_M$ scaled by $\xi_V$.
The horizontal thin solid line is unity, i.e., $\xi_V$ with the scaled error bars.
For understanding the impact of $\zprog$, we show the halo occupation numbers, the average number of galaxies in a halo, $\langle N_{\rm gal}\rangle$ at $z=0$ with varying $\zprog$ as a function of the virial mass of host central subhalos $M_{\rm vir}$ 
in Fig. \ref{fig:hod_Mzprog_z0p0}.
We found a very similar $\zprog$-dependence for  two other samples constructed with  lower number densities of $\ngal=10^{-3}$ and $10^{-4}~h^3~{\rm Mpc}^{-3}$ at $z=0$, as well as for the samples with the same three $\ngal$ values at $z=0.5$ and $1$.

It is naively expected that taking a  higher-$\zprog$ amplifies $\xi_M$ because most massive subhalos at higher-$z$ formed in more biased regions.
However, we observe an `up-and-down' trend in the amplitudes of $\xi_M$ for decreasing redshift.
That is, it rises from $\zprog \simeq 8$ to $\zprog=3\textrm{--}4$ and then turns downward to $\zprog\simeq 0$.
The \smrv{upward} trend from $\smrv{\zprog}=7.77$ to $\smrv{\zprog} \simeq 3\textrm{--}4$ is mainly due to the larger variations in the future MAHs that higher-redshift subhalos will undergo: the descendants of the most massive subhalos at very high-$z$ can be not only well-grown high-mass subhalos but also less-grown low-mass ones at low redshifts.
This is clearly seen in Fig. \ref{fig:hod_Mzprog_z0p0} with $\langle N_{\rm gal}\rangle$ of $\zprog=7.77$.
As a consequence of more low-mass subhalos in the sample, the amplitude of $\xi_M$ is suppressed.
This particularly decreases the one-halo term at $r\lesssim1~\hiMpc$ of $\xi_M$ as Fig. \ref{fig:xi_diff_z0p0} clearly shows the up trend is more prominent at smaller-$r$.
This is because the number of central-satellite pairs in a halo is decreased.

We next discuss the \smrv{downward} trend.
Fig. \ref{fig:xi_diff_z0p0} shows that the overall amplitude of $\xi_M$ peaks at $\zprog\simeq3\textrm{--}4$.
$\xi_M$ with a lower-$\zprog$ is more suppressed, especially in the one-halo term range.
This is due to the mass stripping of satellite subhalos during accretion onto their host halos \citep[e.g.,][]{Reddick13}.
Rank\smrv{-}ordering using subhalo mass near the observation time loses satellite subhalos in the resultant sample.
This is reflected in Fig. \ref{fig:hod_Mzprog_z0p0} which clearly shows that $\langle N_{\rm gal}\rangle$ with lower-$\zprog$ are more suppressed at the high-mass range due to the loss of satellites.
We found that the $\zprog$ values at the transition of up and down trend of $\xi_M$ for the sample with $n_{\rm gal}=10^{-4}~h^3~{\rm Mpc}^{-3}$ is $\simeq0.6$ and lower than the sample with $n_{\rm gal}=10^{-2}~h^3~{\rm Mpc}^{-3}$.
This is because the satellite fraction of the low-$\ngal$ threshold samples is intrinsically low, and then the impact coming from satellite subhalos becomes relatively small.
Hence the overall amplitude of $\xi_M$ can keep high for low $\zprog$.

It is known that the $\vpeak$ model reproduces observed galaxy clustering \smrv{at least for some types of galaxies}.
As shown in Fig. \ref{fig:xi_diff_z0p0}, the 2PCFs predicted by the $\Mprog$ model are similar to or even more amplified than those by the $\vpeak$ model.
It is also known that introducing a scatter between the subhalo and the galaxy properties decreases clustering amplitudes.
Therefore, by varying $\zprog$ and $\sigma_M$, the $\Mprog$ model is expected to  reproduce the observed clustering.
In Appendix \ref{sec:appendix}, we discuss the best matching $\xi_M$ and $\xi_V$ for the  subhalo samples constructed with different number density threshold ($\ngal=10^{-2},~10^{-3}$ and $10^{-4}~h^3~{\rm Mpc}^{-3}$) at $z=0,~0.5$ and $1$.

\subsection{\smrv{Observations}}
\label{sec:obs}
\begin{table}
\centering
\caption{Summary of the incompleteness-corrected number density \smrv{$n_{\rm gal}$, the integral constraint ${\rm IC}$, and the bias factor $b$ for the linear bias model} of each stellar mass threshold sample at $0.30\leq z<0.55$ in \citet{Ishikawa20}. The threshold mass is in units of $\log_{10}[M_{*,~{\rm lim}}/(h^{-2}M_\odot)]$.}
\label{tab:obs_ngal}
\begin{tabular}{cccc}
\hline
threshold mass & $10^3\ngal~[h^3~{\rm Mpc}^{-3}]$ & $10^3{\rm IC}$ & $b$\\
\hline
$11.0$ & $0.175$ & $1.53$ & $1.82$\\
$10.8$ & $0.596$ & $1.27$ & $1.90$\\
$10.6$ & $1.458$ & $1.27$ & $1.75$\\
$10.4$ & $2.689$ & $1.27$ & $1.60$\\
$10.2$ & $4.838$ & $1.26$ & $1.49$\\
$10.0$ & $7.799$ & $1.24$ & $1.43$\\
$9.8$ & $10.42$ & $1.21$ & $1.38$\\
$9.6$ & $12.68$ & $1.20$ & $1.33$\\
$9.4$ & $16.14$ & $1.18$ & $1.30$\\
$9.2$ & $19.50$ & $1.18$ & $1.26$\\
$9.0$ & $23.36$ & $1.18$ & $1.23$\\
$8.8$ & $27.03$ & $1.17$ & $1.18$\\
$8.6$ & $30.74$ & $1.18$ & $1.13$\\
\hline
\end{tabular}
\end{table}
\smrv{
In this paper, we fit the predictions of both $\Mprog$ and $\vpeak$ models to the observed ACFs of the photo-$z$ galaxies at $0.30\leq z<0.55$ from the Subaru HSC survey.
\cite{Ishikawa20} reported the ACFs for the 13 stellar mass threshold samples with the lower limits from $\log_{10}[M_{*,~{\rm lim}}/(h^{-2}M_\odot)]=11$ to $8.6$ with a bin size of $0.2$ dex.
Their observations are attractive because they measured the ACFs in a stellar mass bin as fine as $0.2$ dex and down to a small \smrv{angular} scale of $\simeq4\times10^{-4}$ deg which corresponds to $\simeq8~\hikpc$ at $z=0.43$\footnote{The redshift of $z=0.43$ is close to the peak of the redshift distributions of the observed
galaxies.}, using data over a large area of 178 deg$^2$ of the HSC survey.
Table \ref{tab:obs_ngal} summarizes the number density of each stellar mass threshold sample for which \cite{Ishikawa20} measured the ACFs.
Note that these values are corrected for incompleteness.
}

\smrv{
We remeasure the ACFs for the same sample and in the same scale range as in \cite{Ishikawa20}, using the galaxy catalogs and the random points used in the paper.
Our measurement is done with more careful treatments in some aspects.
We use the {\sc Corrfunc} package to measure the ACFs by the Landy-Szalay estimator \citep{LS93}
\begin{align}
    \omega(\theta)=\frac{{\rm DD(\theta)}-2{\rm DR(\theta)}+{\rm RR(\theta)}}{\rm RR(\theta)}
\end{align}
where $\theta$ is the separation angle on the sky and ${\rm DD,~DR,~RR}$ is the normalized count of the galaxy-galaxy, galaxy-random and random-random pairs, respectively.
}

\smrv{
We estimate the covariance matrix $C_{{\rm obs},~ij}$ of the ACFs by the `omit-one' jackknife resampling \citep{Norberg09} as 
\begin{align}
    C_{{\rm obs},~ij}=\frac{N_{\rm sub}-1}{N_{\rm sub}}\sum_{n=1}^{N_{\rm sub}}
    \left[\omega_{n}(\theta_i)-\bar\omega(\theta_i)\right]
    \left[\omega_{n}(\theta_j)-\bar\omega(\theta_j)\right]
\end{align}
where $N_{\rm sub}$ is the number of subfields, $\omega_n$ is the ACF for the $n$-th jackknife realization, and $\bar\omega$ is the average of $\omega_n$ as $\bar\omega(\theta)=\sum_{n=1}^{N_{\rm sub}}\omega_n(\theta)/N_{\rm sub}$.
We divide the survey area, consisting of six fields, into $N_{\rm sub}=150$ subfields by the $k$-means algorithm\footnote{\url{https://github.com/esheldon/kmeans_radec}} as employed by \cite{Okumura21}.
This allows for division by non-artificial shapes of subfields while \cite{Ishikawa20} employed the rectangular-shaped subfield.
}

\smrv{
We correct the measured ACFs for unavoidable systematic underestimation due to the finite size of the survey area, known as the integral constraint ${\rm IC}$ \citep{Peebles76}, assuming the linear bias model.
The measured ACFs $\omega_{\rm measured}$ is related to the corrected ACFs $\omega_{\rm corrected}$ as
\begin{align}
    \omega_{\rm corrected}(\theta)
    =\omega_{\rm measured}(\theta)+b^2{\rm IC}
    =b^2\omega_{\rm nl}(\theta)
\end{align}
where $b$ is the linear bias, $\omega_{\rm nl}$ is the non-linear ACF of matter.
We obtain $\omega_{\rm nl}$ by projecting the real-space non-linear matter correlation function $\xi_{\rm nl}$ as a function of the comoving distance $r$ using the Limber approximation \citep{Limber53,Simon07} as
\begin{align}
    \omega_{\rm nl}(\theta)&=2\int_0^\infty \rd z~\frac{p^2(z)}{\rd \chi/\rd z} \int_0^\infty \rd u ~\xi_{\rm nl}\left(r= \sqrt{u^2+\chi^2(z)\theta^2}\right)\\
    &=2\int_0^\infty \rd z~\frac{p^2(z)}{\rd \chi/\rd z} \int_{\chi(z)\theta}^\infty\rd r \frac{r\xi_{\rm nl}(r)}{\sqrt{r^2-\chi^2(z)\theta^2}},
    \label{eq:limber}
\end{align}
where $p(z)$ is the normalized redshift distribution of the observed galaxies measured by \cite{Ishikawa20}, $\chi(z)$ is the comoving distance to the redshift $z$ and $u$ is the comoving distance along the line-of-sight.
We compute $\xi_{\rm nl}$ at $z=0.43$ for the adopted cosmology with the revised {\sc Halofit} fitting formula \citep{Takahashi12} implemented in the {\sc CLASS} code\footnote{\url{https://github.com/nickhand/classylss}} \citep{class1} available through {\sc nbodykit}.
The value of ${\rm IC}$ is given by \citep{Roche99}
\begin{align}
    {\rm IC}=\frac{\sum_i \omega_{\rm nl}(\theta_i){\rm RR}(\theta_i)}{\sum_i {\rm RR}(\theta_i)}.
\end{align}
We sum up to $\theta = 15~{\rm deg}$, which roughly equals to the size of the six observation fields.
We found that the values of ${\rm IC}$ range from $0.0015$ to $0.0012$ for the all samples.
Note that the value of {\rm IC} varies from sample to sample because $p(z)$ is different.
We search the value of $b$ which gives the least chi-square computed as
\begin{align}
    &\chi_0^2=\sum_{i,j}\Delta_{0i}C^{-1}_{{\rm obs},~ij}\Delta_{0j},\\
    &\Delta_{0i}=\omega_{\rm measured}(\theta_i)-b^2[\omega_{\rm nl}(\theta_i)-{\rm IC}].
\end{align}
We fit in the limited angle range of $0.4<\theta~[{\rm deg}]<1.6$, corresponding to the range from $8~\hiMpc$ to $32~\hiMpc$, where the linear bias model can be valid \citep{Sugiyama23}.
The values of ${\rm IC}$ and $b$ used hereafter are listed in Table \ref{tab:obs_ngal}.
}

\subsection{Fitting to observed clustering}
\label{sec:how_to_fit}
\smrv{For the implementation of the two SHAM models, we use the halo/subhalo catalog from the mini- and Shin-Uchuu simulations at $z=0.43$, 
which is close to the peak of the redshift  distributions of the observed galaxies.}
In fitting the predictions of the $\Mprog$ model to observed clustering, as we stated, we treat the two parameters $\zprog$ and $\sigma_M$ as free parameters.
Specifically, we take 40 values for $\zprog$ from $0.49$ to $13.93$ \smrv{for the mini run and 60 values from $0.49$ to $19.96$ for the Shin run}, and 20 values for $\sigma_M$ from $0$ to $0.19$ with a linear spacing of $0.01$.
As discussed in Sec. \ref{sec:motivation}, we allow for a stellar mass dependence of $\zprog$.
As well as $\zprog$, we treat $\sigma_M$ as a function of the galaxy stellar mass.

We also use the $\vpeak$ model for comparison.
We perturb $\vpeak$ by the same method as in the $\Mprog$ model to account for the scatter between $\vpeak$ and $M_*$ as
\begin{align}
    \log_{10}V_{\rm pert} = [1+\mathcal{N}(0,~\sigma_V)]\log_{10}\vpeak,
\end{align}
where $\sigma_V$ is the standard deviation of the Gaussian distribution $\mathcal{N}$ with the zero-mean.
$\sigma_V$ is the only free parameter \smrv{of the $\vpeak$ model} and taken to be from $0$ to $0.49$ with the linear spacing of $0.01$.
We also treat this parameter as a function of the galaxy stellar mass.

Allowing all free parameters in both models to depend on stellar mass, we construct the subhalo samples  corresponding to the stellar mass threshold galaxy samples in a self-consistent manner as follows.
First, we assume that the free parameters are constant for the most massive sample, i.e., the sample of $\log_{10}[M_*/(h^{-2}M_\odot)]\geq11$ in this paper.
For this sample, we simply perform the rank-ordering SHAM using the perturbed $\Mprog$ or $\vpeak$.
We measure the ACFs for each parameter set and find the best-fit set.
Then we temporally exclude the subhalos which are assigned with the sample galaxies by the best-fit parameter set from the whole subhalo catalog.
Next, we abundance-match using the rest of the subhalos for the galaxy sample of $10.8\leq\log_{10}[M_*/(h^{-2}M_\odot)]<11$ assuming that the free parameters are constant in this narrow range of the stellar mass.
The number density of subhalos in this bin given by the difference between the two samples of $\log_{10}[M_*/(h^{-2}M_\odot)]\geq11$ and $\log_{10}[M_*/(h^{-2}M_\odot)]\geq10.8$.
Combining the subhalos assigned with galaxies of $\log_{10}[M_*/(h^{-2}M_\odot)]\geq11$ by the best-fit parameter set and the subhalos assigned with galaxies of $10.8\leq\log_{10}[M_*/(h^{-2}M_\odot)]<11$ by each parameter set, we obtain the subhalo sample for the galaxies with $\log_{10}[M_*/(h^{-2}M_\odot)]\geq10.8$.
By comparing the predicted ACFs with the observation for the galaxies with $\log_{10}[M_*/(h^{-2}M_\odot)]\geq10.8$, we obtain the best-fit parameter set for the galaxies of $10.8\leq\log_{10}[M_*/(h^{-2}M_\odot)]<11$.
We repeat this procedure every $0.2$ dex bin until reaching the galaxy sample with $\log_{10}[M_*/(h^{-2}M_\odot)]\geq8.6$.

To compute the model predictions of ACFs \smrv{$\omega_{\rm model}(\theta)$}, we first measure the real-space 2PCFs \smrv{$\xi_{\rm model}(r)$} for the subhalo samples constructed with the $\Mprog$ and $\vpeak$ models.
Then we project \smrv{$\xi_{\rm model}$} along the line of sight to obtain \smrv{$\omega_{\rm model}$} using the Limber approximation, \smrv{i.e., replacing $\xi_{\rm nl}$ and $\omega_{\rm nl}$ in Eq.(\ref{eq:limber}) with $\xi_{\rm model}$ and $\omega_{\rm model}$, respectively.} 
In doing so, the evolution of $\xi_{\rm model}$ over the redshift range $0.30\leq z<0.55$ is ignored.
\smrv{To be consistent with $\omega_{\rm measured}$, we subtract the correction term for the integral constraint $b^2{\rm IC}$ from $\omega_{\rm model}$ as
\begin{align}
    \omega_{\rm model}(\theta)\rightarrow\omega_{\rm model}(\theta)-b^2{\rm IC}.
\end{align}}

\smrv{To constrain the parameters for each threshold sample, we calculate chi-square values at every grid point in the parameter space as follows:
\begin{align}
    &\chi^2=\sum^{N_{\rm bin}}_{i,j}\Delta_iC^{-1}_{ij}\Delta_j,\\
    &\Delta_i=\omega_{\rm measured}(\theta_i)-\omega_{\rm model}(\theta_i).
\end{align}
$N_{\rm bin}=19$ is the number of the $\theta$ bin.
$C^{-1}_{ij}$ is inverse of the covariance matrix evaluated as
\begin{align}
    C_{ij}=C_{{\rm obs},~ij}/f_{\rm Hartlap}+C_{{\rm model},~ij}
\end{align}
where $f_{\rm Hartlap}=(N_{\rm sub}-N_{\rm bin}-2)/(N_{\rm sub}-1)$ is the Hartlap factor for accounting for the finite number of jackknife realizations \citep{Hartlap07}.
As well as the observations discussed above, we estimate the covariance matrix for $\omega_{\rm model}$, $C_{{\rm model},~ij}$, using the jackknife resampling by dividing the whole simulation volume into $27$ subvolumes.
We hereafter call the parameter set which gives the minimum $\chi^2$ the best-fit set.
To estimate the $1\sigma$ range of the parameters, we compute the likelihood as $L\propto\exp\left(-\chi^2/2\right)$.
The $1\sigma$ range is defined as the range that contains the parameter with the maximum likelihood and where the integral of the likelihood is $0.68$.
}

\section{Results \smrv{and discussions}}
\label{sec:result}
Here we present the results of fitting to the observed ACFs and show that the $\Mprog$ model matches with the observations better than the $\vpeak$ model.
We discuss the \smrv{constraints on the} free parameters in the two models.
Finally, we study the inferred satellite fractions.

\subsection{The ACFs $\omega(\theta)$ at $z\simeq0.4$: the observation versus the SHAM models}
\begin{figure*}
\includegraphics[width=2\columnwidth]{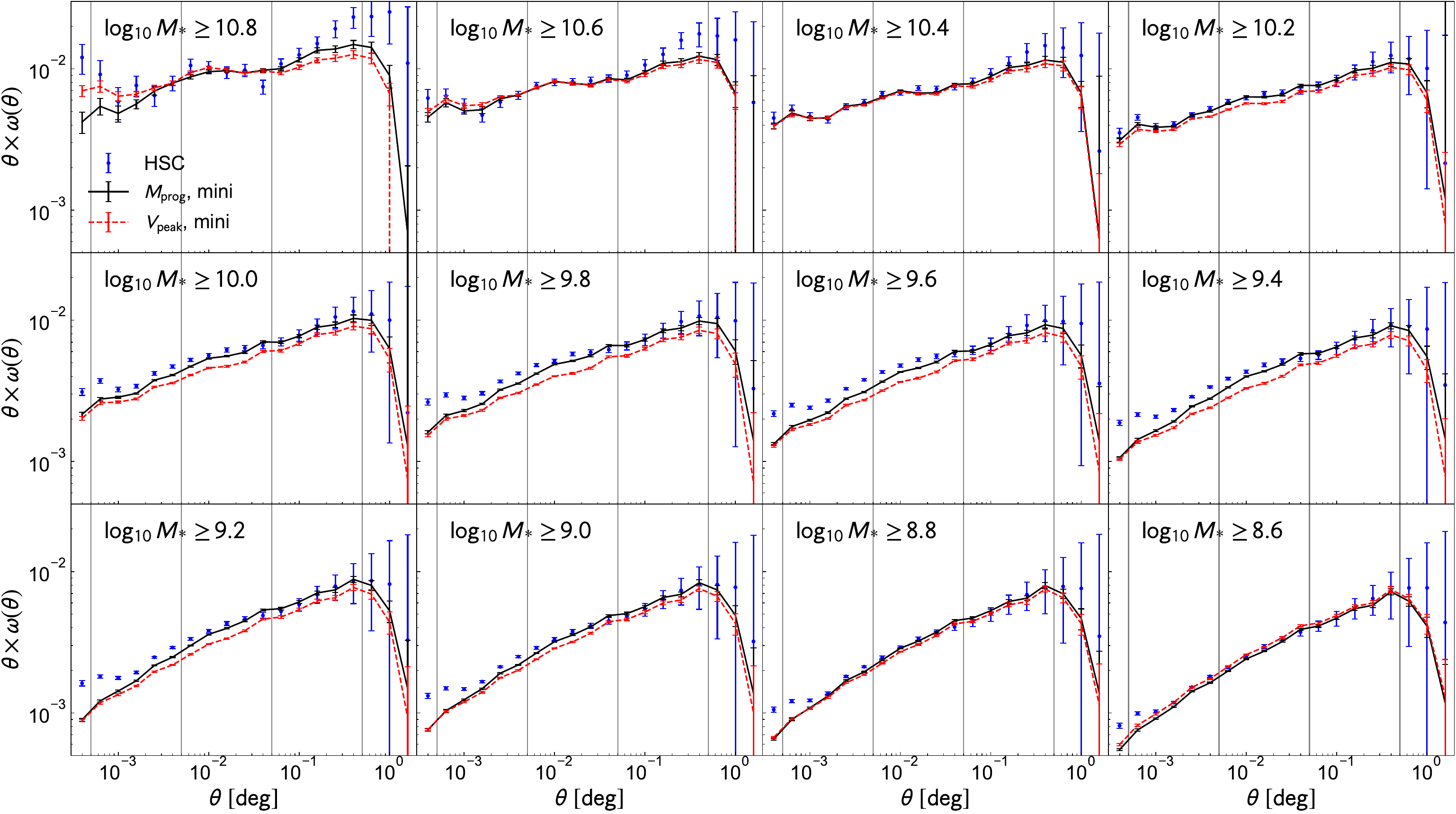}
\includegraphics[width=2\columnwidth]{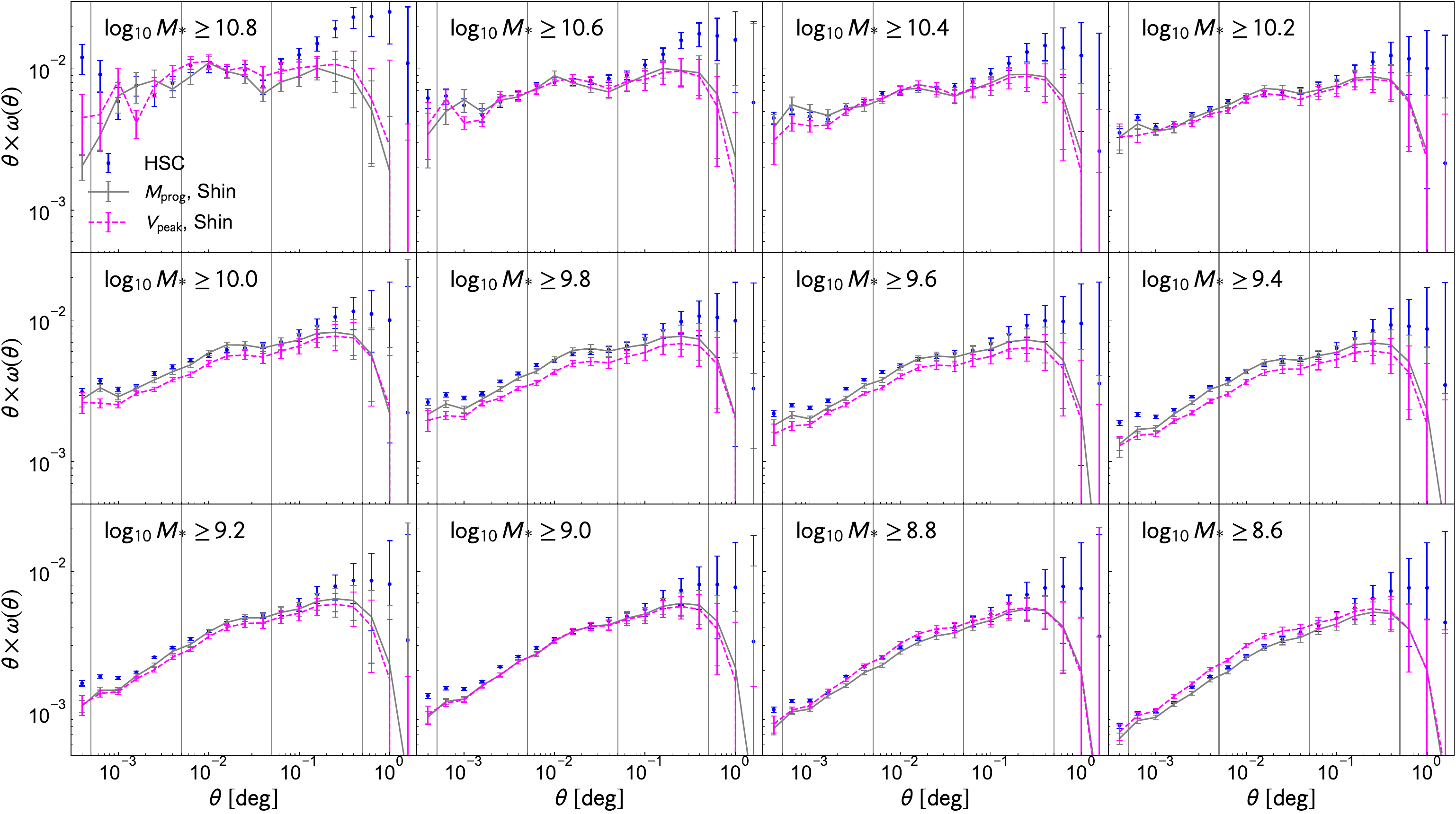}
\caption{
The ACFs $\omega(\theta)$ at $z\simeq0.4$ of the stellar mass threshold samples.
The mass ranges are noted in each panel, where the stellar mass $M_*$ is in units of $h^{-2}M_\odot$.
The blue data points with  error bars are the observational results \smrv{from the Subaru HSC survey (see Sec.~\ref{sec:obs})}.
\smrv{The top and bottom panels present the best-fit results in the mini- and Shin-Uchuu simulations, respectively.
The solid and dashed lines show the results of the $\Mprog$ and $\vpeak$ models, respectively.}
The vertical gray solid lines are the angles corresponding to  comoving scales of $0.01,~0.1,~1$, and $10~\hiMpc$ at $z=0.43$, respectively.
}
\label{fig:w_results}
\end{figure*}
\begin{figure}
\includegraphics[width=\columnwidth]{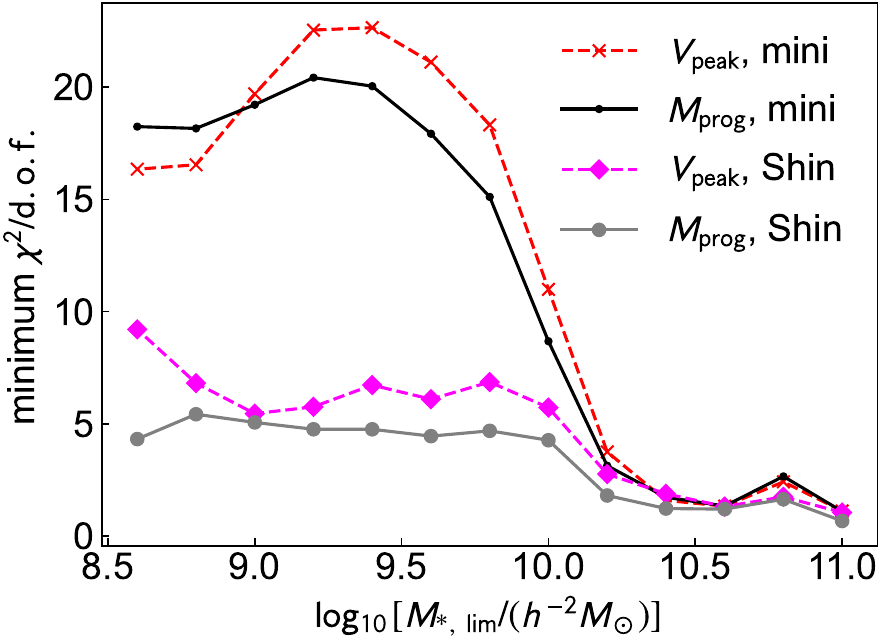}
\caption{
\smrv{The reduced-$\chi^2$ values which are given by the best-fit parameter sets as a function of the threshold stellar mass $\log_{10}[M_{*,~{\rm lim}}/(h^{-2}M_\odot)]$.
The solid and dashed lines show the results of the $\Mprog$ and $\vpeak$ models, respectively.
The labels in the legend ``mini'' and ``Shin'' means the results from the mini- and Shin-Uchuu simulations, respectively.}
}
\label{fig:chi2min}
\end{figure}
Fig.~\ref{fig:w_results} compares the observed ACFs and the best-fit predictions from the $\Mprog$ and $\vpeak$ models \smrv{in the mini- and Shin-Uchuu simulations}.
For clarity, we show $\theta\times\omega(\theta)$ as the vertical axis rather than $\omega(\theta)$.
The vertical gray thin lines are the angles corresponding to the comoving distances of $0.01,~0.1,~1$ and $10~\hiMpc$ at $z=0.43$.
\smrv{The blue data points with  error bars are the ACFs $\omega_{\rm measured}$ of the HSC galaxies, where $\sqrt{C_{{\rm obs},ii}}$ is quoted as the error bars for $\omega_{\rm measured}(\theta_i)$ (see Sec.~\ref{sec:obs}).}
This figure omits the ACFs for the sample with $\log_{10}[M_*/(h^{-2}M_\odot)]\geq11$ due to the  large error bars.
\smrv{The top and bottom panels present the best-fit $\omega_{\rm model}$ in the mini- and Shin-Uchuu simulations after subtracting the correction term $b^2{\rm IC}$, respectively.
We quote $\sqrt{C_{{\rm model},ii}}$ as the error bars for $\omega_{\rm model}(\theta_i)$.
In each panel, the solid and dashed lines show the results of the $\Mprog$ and $\vpeak$ models, respectively.}

\smrv{
We first discuss the general differences between the mini- and Shin-Uchuu simulations.
For all samples and both models, the ACFs from the Shin run are more suppressed than those from the mini run at larger scales of $\gtrsim10~\hiMpc$, and the ratios of the amplitudes become even $0.3$ at the largest scales.
This is simply because the number of large-scale galaxy pairs would be smaller within the smaller simulation box.
On the other hand, at smaller scales of $\lesssim1~\hiMpc$, we observe that the ACFs from the Shin run are $5\textrm{--}10\%$ more enhanced than those from the mini run, especially for the samples with $\log_{10}[M_{*,~{\rm lim}}/(h^{-2}M_\odot)]\leq10$.
This should be due to the resolution effect as the Shin run can resolve more small satellite subhalos than the mini run.
It is also observed that such enhancement in the $\Mprog$ model is weaker than in the $\vpeak$ model for the samples with $\log_{10}[M_{*,~{\rm lim}}/(h^{-2}M_\odot)]\leq9$.
Thus our $\Mprog$ model has the advantage of being less sensitive to resolution than the $\vpeak$ model.
}

\smrv{
We next compare the $\Mprog$ and $\vpeak$ models.
For this, we show the reduced-$\chi^2$ values which are given by the best-fit parameter sets as a function of the threshold stellar mass $\log_{10}[M_{*,~{\rm lim}}/(h^{-2}M_\odot)]$ in Fig.~\ref{fig:chi2min}.
For the four most massive samples, i.e., the samples with the threshold mass of $\log_{10}[M_{*,~{\rm lim}}/(h^{-2}M_\odot)]\geq10.4$, we see that the best-fit ACFs of both models in both simulations agree with the observations fairly well as the reduced-$\chi^2$ values are $1\textrm{--}2.5$.
This is consistent with the fact that the $\vpeak$ model works well for clustering of the massive galaxies at $z\simeq0.5$ so-called CMASS \citep{Nuza13,RT16,Saito16}.
}

\smrv{
The situation changes for the samples with $9.2\leq\log_{10}[M_{*,~{\rm lim}}/(h^{-2}M_\odot)]\leq10.2$.
The $\Mprog$ model predicts ACFs with higher amplitudes and agrees better with the observation than the $\vpeak$ model at $\lesssim 1~\hiMpc$.
Fig.~\ref{fig:chi2min} shows that the best-fit reduced-$\chi^2$ values from the $\Mprog$ model are lower than the $\vpeak$ model in each simulation.
This is due to the higher satellite fractions of the $\Mprog$ model (see Sec.~\ref{sec:fsat}).
The $\vpeak$ model has been tested against the Sloan Digital Sky Survey main galaxies at $z\simeq0$ and massive galaxies at $z\simeq0.3\textrm{--}0.5$.
Hence the lower-than-observed clustering amplitudes for non-massive galaxies at $z\simeq0.4$ induced by the $\vpeak$ model we found here are not inconsistent with the literature and rather a new shortcoming of the model.
The agreements of the $\Mprog$ model with the observations differ for the simulations.
The $\Mprog$ model in the Shin run gives the higher clustering amplitudes and agrees better with the observation than that in the mini run at $\lesssim1~\hiMpc$.
This is due to the resolution effects, as we stated above.
At the smallest scales of $\lesssim30~\hikpc$, even the $\Mprog$ model in the Shin run underpredicts the observed clustering amplitudes by $10\textrm{--}20\%$.
This implies that our $\Mprog$ model faces its resolution limit and/or needs fine-tuning for more accurate modeling of galaxy-subhalo connections.
The $\Mprog$ model in the mini run appears to be more consistent with observation than the $\vpeak$ model in the Shin run, but the chi-square value is larger for the former than the latter because the former's $C_{\rm model}$ is smaller.
}

\smrv{
For the samples with $\log_{10}[M_{*,~{\rm lim}}/(h^{-2}M_\odot)]\leq9$, the $\Mprog$ model in the Shin run still provides the best agreements with the observation at $\lesssim 1~\hiMpc$ although its underprediction of the clustering amplitudes at $\lesssim30~\hikpc$.
The difference from the other samples is that the $\vpeak$ model is not necessarily lower in amplitude than the $\Mprog$ model in each simulation.
For the least massive sample of $\log_{10}[M_{*,~{\rm lim}}/(h^{-2}M_\odot)]=8.6$, the $\vpeak$ model has higher amplitude than the $\Mprog$ model in each simulation.
This can also be explained by the high satellite fraction (see Sec~\ref{sec:fsat}).
}

\smrv{
In short, compared to the widely-used $\vpeak$ model, our $\Mprog$ model has a higher capability to explain the observed clustering signal of non-massive galaxies at $z\simeq0.4$.
}

\subsection{The \smrv{parameter constraints}}
\begin{table*}
\centering
\caption{The best-fit values of parameters \smrv{and the $1\sigma$ range from the likelihood analysis} for the stellar mass bin samples in the $\Mprog$ model ($\sigma_M,~\zprog$) and the $\vpeak$ model ($\sigma_V$). The mass range is in units of $\log_{10}[M_*/(h^{-2}M_\odot)]$. \smrv{The results from the mini- and Shin-Uchuu simulations are listed separately.}}
\label{tab:best_params}
\begin{tabular}{c c ccc c ccc}
\hline
~ & ~ & ~ & mini-Uchuu & ~ & ~ & ~ & Shin-Uchuu & ~\\
mass range & ~ & $\sigma_M$ & $\zprog$ & $\sigma_V$ & ~ & $\sigma_M$ & $\zprog$ & $\sigma_V$\\
\hline
$>11$ & ~~~~~ & $0.09,~~~0.09^{+0.05}_{-0.01}$ & $1.65,~~~9.48^{+0.0}_{-8.99}$ & $0.16,~~~0.16^{+0.07}_{-0.02}$ & ~~~~~ & $0.08,~~~0.07^{+0.05}_{-0.03}$ & $8.58,~~~8.58^{+7.66}_{-1.06}$ & $0.19,~~~0.19^{+0.01}_{-0.02}$\\
$10.8\textrm{--}11.0$     & ~~~~~     & $0.03,~~~0.04^{+0.02}_{-0.01}$     & $0.49,~~~0.49^{+0.73}_{-0.0}$     & $0.11,~~~0.11^{+0.03}_{-0.02}$     & ~~~~~     & $0.06,~~~0.03^{+0.03}_{-0.02}$     & $0.78,~~~0.78^{+13.54}_{-0.29}$     & $0.13,~~~0.13^{+0.02}_{-0.03}$\\
$10.6\textrm{--}10.8$     & ~~~~~     & $0.05,~~~0.05^{+0.02}_{-0.02}$     & $1.22,~~~1.22^{+0.81}_{-0.36}$     & $0.09,~~~0.09^{+0.03}_{-0.0}$     & ~~~~~     & $0.02,~~~0.02^{+0.04}_{-0.02}$     & $0.56,~~~0.56^{+0.66}_{-0.0}$     & $0.11,~~~0.11^{+0.07}_{-0.01}$\\
$10.4\textrm{--}10.6$     & ~~~~~     & $0.02,~~~0.03^{+0.01}_{-0.03}$     & $1.03,~~~1.12^{+0.77}_{-0.18}$     & $0.07,~~~0.07^{+0.02}_{-0.01}$     & ~~~~~     & $0.03,~~~0.03^{+0.02}_{-0.01}$     & $1.12,~~~1.12^{+0.1}_{-0.09}$     & $0.1,~~~0.1^{+0.0}_{-0.05}$\\
$10.2\textrm{--}10.4$     & ~~~~~     & $0.0,~~~0.0^{+0.01}_{-0.0}$     & $1.77,~~~1.65^{+0.51}_{-0.12}$     & $0.01,~~~0.01^{+0.02}_{-0.01}$     & ~~~~~     & $0.04,~~~0.04^{+0.01}_{-0.01}$     & $2.46,~~~2.46^{+0.49}_{-0.15}$     & $0.03,~~~0.03^{+0.01}_{-0.02}$\\
$10.0\textrm{--}10.2$     & ~~~~~     & $0.0,~~~0.0^{+0.01}_{-0.0}$     & $2.31,~~~2.31^{+0.47}_{-0.14}$     & $0.02,~~~0.02^{+0.02}_{-0.02}$     & ~~~~~     & $0.02,~~~0.02^{+0.01}_{-0.01}$     & $2.46,~~~2.46^{+4.11}_{-0.81}$     & $0.02,~~~0.02^{+0.01}_{-0.02}$\\
$9.8\textrm{--}10.0$     & ~~~~~     & $0.0,~~~0.0^{+0.01}_{-0.0}$     & $3.31,~~~3.31^{+0.95}_{-0.36}$     & $0.02,~~~0.02^{+0.01}_{-0.02}$     & ~~~~~     & $0.03,~~~0.03^{+0.16}_{-0.0}$     & $2.78,~~~2.78^{+7.42}_{-1.46}$     & $0.1,~~~0.1^{+0.19}_{-0.07}$\\
$9.6\textrm{--}9.8$     & ~~~~~     & $0.0,~~~0.0^{+0.01}_{-0.0}$     & $2.17,~~~2.17^{+1.76}_{-0.14}$     & $0.0,~~~0.0^{+0.02}_{-0.0}$     & ~~~~~     & $0.01,~~~0.19^{+0.0}_{-0.18}$     & $1.43,~~~1.43^{+11.74}_{-0.1}$     & $0.16,~~~0.16^{+0.33}_{-0.08}$\\
$9.4\textrm{--}9.6$     & ~~~~~     & $0.0,~~~0.0^{+0.01}_{-0.0}$     & $3.93,~~~3.93^{+0.34}_{-0.8}$     & $0.0,~~~0.0^{+0.02}_{-0.0}$     & ~~~~~     & $0.16,~~~0.16^{+0.03}_{-0.01}$     & $16.92,~~~16.92^{+1.46}_{-4.3}$     & $0.09,~~~0.09^{+0.27}_{-0.02}$\\
$9.2\textrm{--}9.4$     & ~~~~~     & $0.0,~~~0.0^{+0.01}_{-0.0}$     & $2.31,~~~2.95^{+0.0}_{-1.06}$     & $0.01,~~~0.01^{+0.02}_{-0.01}$     & ~~~~~     & $0.03,~~~0.02^{+0.02}_{-0.01}$     & $1.22,~~~1.22^{+0.21}_{-0.28}$     & $0.06,~~~0.06^{+0.06}_{-0.04}$\\
$9.0\textrm{--}9.2$     & ~~~~~     & $0.0,~~~0.0^{+0.01}_{-0.0}$     & $1.9,~~~1.9^{+1.06}_{-0.24}$     & $0.01,~~~0.01^{+0.01}_{-0.01}$     & ~~~~~     & $0.02,~~~0.02^{+0.04}_{-0.01}$     & $1.03,~~~0.94^{+0.59}_{-0.08}$     & $0.26,~~~0.26^{+0.05}_{-0.16}$\\
$8.8\textrm{--}9.0$     & ~~~~~     & $0.01,~~~0.0^{+0.02}_{-0.0}$     & $1.54,~~~1.54^{+0.36}_{-0.11}$     & $0.02,~~~0.02^{+0.01}_{-0.02}$     & ~~~~~     & $0.01,~~~0.01^{+0.08}_{-0.01}$     & $0.56,~~~0.56^{+0.66}_{-0.07}$     & $0.31,~~~0.31^{+0.07}_{-0.19}$\\
$8.6\textrm{--}8.8$     & ~~~~~     & $0.0,~~~0.0^{+0.02}_{-0.0}$     & $0.49,~~~0.56^{+0.22}_{-0.07}$     & $0.04,~~~0.04^{+0.18}_{-0.01}$     & ~~~~~     & $0.12,~~~0.12^{+0.01}_{-0.1}$     & $0.86,~~~0.86^{+0.26}_{-0.23}$     & $0.47,~~~0.47^{+0.02}_{-0.4}$\\
\hline
\end{tabular}
\end{table*}
\begin{figure}
\includegraphics[width=\columnwidth]{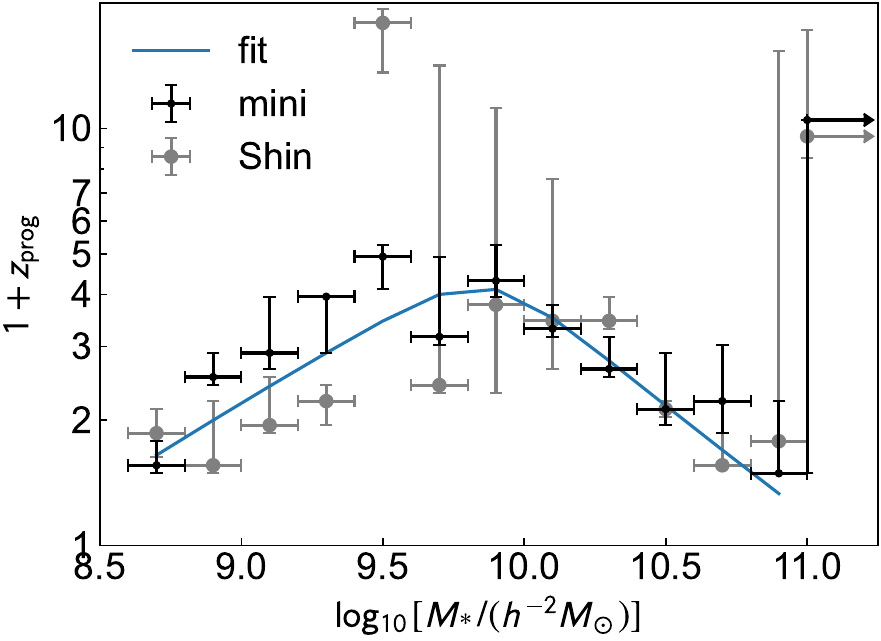}
    \caption{
    \smrv{The $1\sigma$ range of the $\zprog$ parameter of the $\Mprog$ model for the stellar mass bin samples in the mini- and Shin-Uchuu simulations.}
    Note that only the highest mass range represents the threshold  sample of $\log_{10}[M_*/(h^{-2}M_\odot)]\geq11$, not a binned one.
    \smrv{The solid line shows the best-fit broken power-law model (see Eqs.\ref{eq:broken} and \ref{eq:broken_param})}.
    }
    \label{fig:best_param_Mprog}
\end{figure}
\smrv{Table \ref{tab:best_params} summarizes the parameter constraints, i.e., the best-fit values and the 1$\sigma$ ranges} of $\sigma_M,~\zprog$ in the $\Mprog$ model and $\sigma_V$ in the $\vpeak$ model for the stellar mass bin samples.
\smrv{Fig.~\ref{fig:best_param_Mprog} shows the 1$\sigma$ range of $\zprog$ as a function of stellar mass.}

The \smrv{obtained} characteristic redshift \smrv{in} the $\Mprog$ model, $\zprog$, displays an interesting \smrv{trend in both simulations}.
As shown in Fig.~\ref{fig:best_param_Mprog}, $\zprog$ appears to peak at \smrv{$9.4<\log_{10}[M_*/(h^{-2}M_\odot)]<10$}.
As discussed in Sec.~\ref{sec:motivation}, this \smrv{behavior} is expected, and related to the two phases of stellar mass growth in galaxies.
In the two-phase \smrv{growth} scenario, $\zprog$ would increase \smrv{for higher mass galaxies when the in-situ star formation dominates, but then turn to decrease for more massive galaxies in which the ex-situ star accretion becomes more important.}
In other words, $\zprog$ as a function of the galaxy stellar mass has a peak at the intermediate mass.
This is indeed seen in Fig. \ref{fig:best_param_Mprog}.
Also, it is suggested that the ex-situ star accretion is efficient for the galaxies with \smrv{$\log_{10}[M_*/(h^{-2}M_\odot)]\gtrsim10$} at $z\simeq0.4$, and the in-situ star formation is dominant in the lower mass galaxies.
\smrv{
The behaviors of $\zprog$ constrained with the two simulations are qualitatively consistent.
We note that, however, there is an outlier at $9.4\leq\log_{10}[M_*/(h^{-2}M_\odot)]\leq9.6$ in the results with the Shin-Uchuu simulation, which is possibly due to cosmic variance in the smaller simulation box.
}

\smrv{
We fit the mass dependence of $\zprog$ with a broken power-law form as
\begin{align}
    1+\zprog(M_*)=&(1+z_{\rm prog,1})\left(\frac{\log_{10}M_*}{\log_{10}M_{\rm tr}}\right)^{a_1}\nonumber\\
    &\times\left[ \frac{1}{2}\left\{1+\left(\frac{\log_{10}M_*}{\log_{10}M_{\rm tr}}\right)^{1/d}\right\} \right]^{(a_2-a_1)d},
    \label{eq:broken}
\end{align}
where masses $M_*$ and $M_{\rm tr}$ are in units of $h^{-2}M_\odot$.
The function has five parameters $z_{\rm prog,1},~M_{\rm tr},~a_1,~a_2$ and $d$: $\zprog$ reaches its peak $z_{\rm prog,1}$ at the transition mass $M_*=M_{\rm tr}$, and $a_1$ and $a_2$ are the slope at $M*<M_{\rm tr}$ and $M*>M_{\rm tr}$, and $d$ controls the width of transition of the power, respectively.
}

\smrv{
We fit Eq.(\ref{eq:broken}) to combined $\zprog$ from the mini- and Shin-Uchuu simulations but not using the point of $\log_{10}[M_*/(h^{-2}M_\odot)]\geq11$.
The best-fit parameters are below
\begin{align}
    &z_{\rm prog,1}=3.14\pm0.60,~\log_{10}[M_{\rm tr}/(h^{-2}M_\odot)]=9.88\pm0.10,\nonumber\\
    &a_1=8.47\pm1.43,~a_2=-13.0\pm3.3,~d=0.0101\pm0.0134.
    \label{eq:broken_param}
\end{align}
We show the best-fit broken power-law model in Fig.~\ref{fig:best_param_Mprog} by the solid line.
Thus the mass at which the transition of stellar mass growth mode occurs at $z\simeq0.4$ is estimated to be $M_{\rm tr}\simeq10^{9.9}h^{-2}M_\odot$.
}

\smrv{
For some samples, as seen in Fig.~\ref{fig:w_results}, the best-fit $\omega_{\rm model}$ lacks clustering amplitudes, especially at smaller scales.
Therefore the scatter parameters $\sigma_M$ and $\sigma_V$ become quite low values or even zero to have higher amplitudes.
The scatter parameters, $\sigma_M$ and $\sigma_V$, in the mini-Uchuu simulation are lower than those in the Shin run for almost all samples simply due to the resolution effect.
}

\subsection{The inferred satellite fraction}
\label{sec:fsat}
\begin{figure}
\includegraphics[width=\columnwidth]{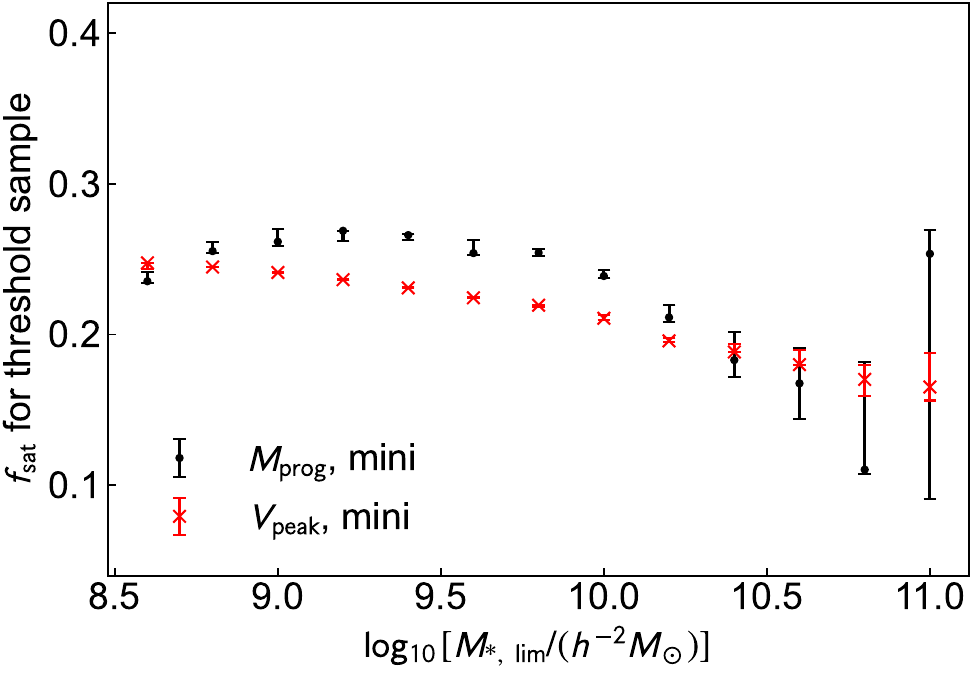}
\includegraphics[width=\columnwidth]{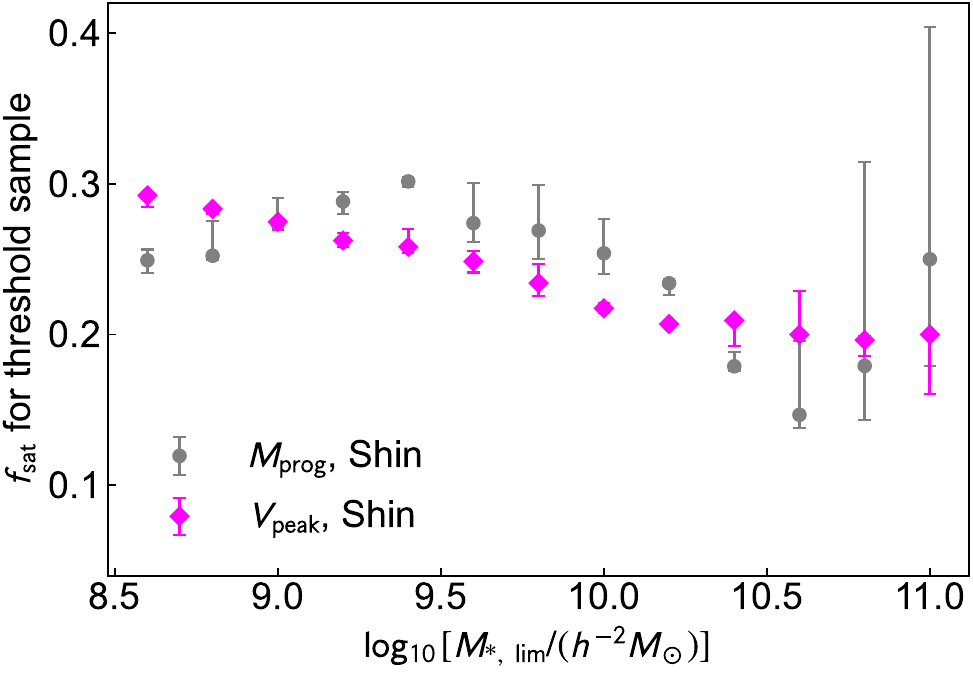}
\caption{
The satellite fraction $f_{\rm sat}$ of the threshold sample as a function of the threshold mass $\log_{10}[M_{*,~{\rm lim}}/(h^{-2}M_\odot)]$ inferred by \smrv{the $\Mprog$ and $\vpeak$ models}.
\smrv{The results from the mini- and Shin-Uchuu simulations are shown in the top and bottom panels, respectively.}
}
\label{fig:fsat_thr_results}
\end{figure}
\begin{figure}
\includegraphics[width=\columnwidth]{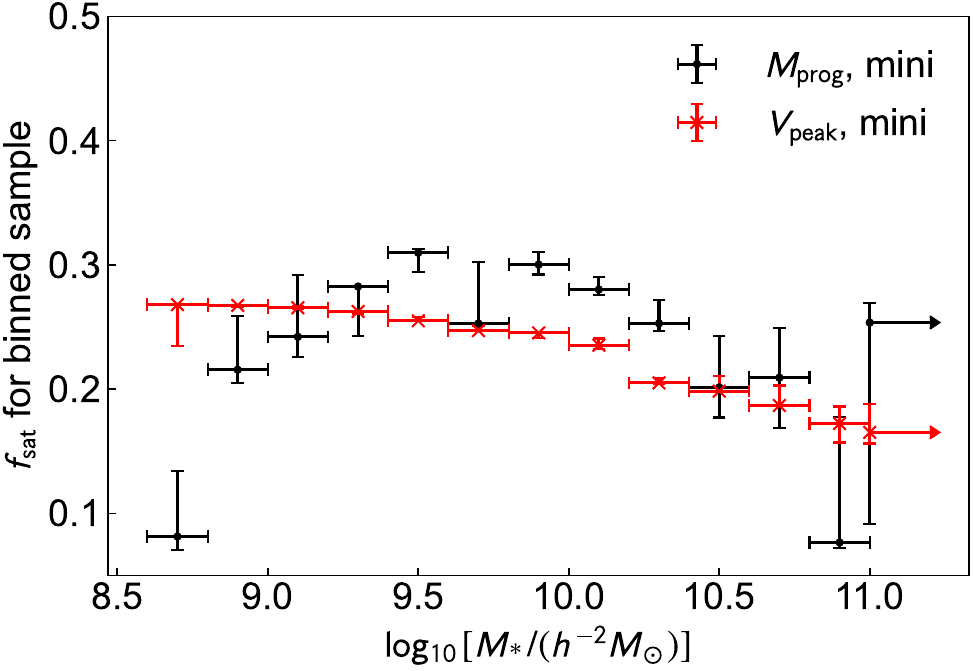}
\includegraphics[width=\columnwidth]{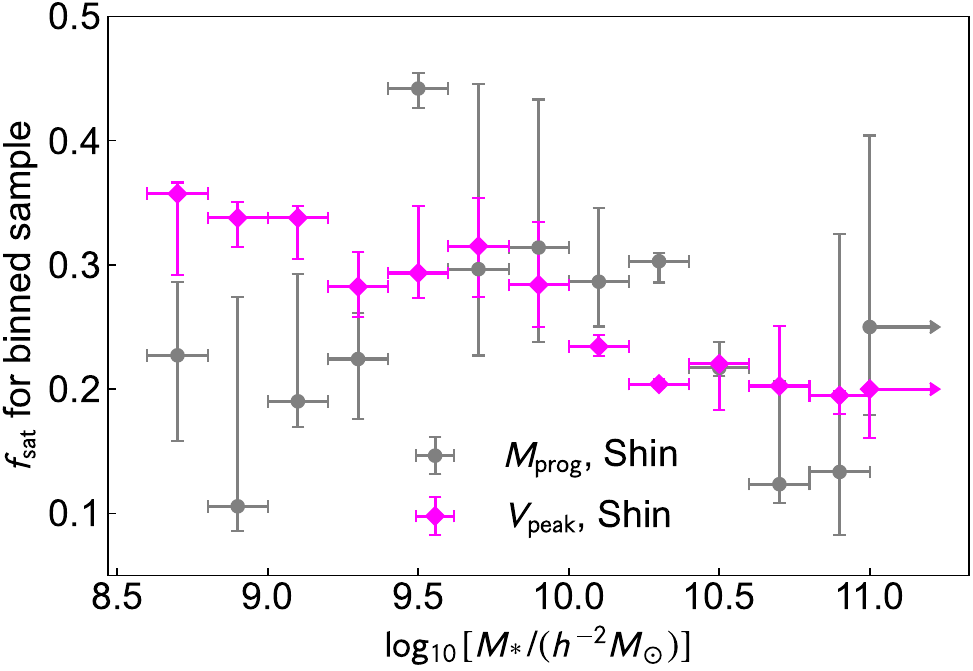}
\caption{
Very similar to Fig. \ref{fig:fsat_thr_results} but the inferred satellite fractions for the stellar mass bin samples.
}
\label{fig:fsat_bin_results}
\end{figure}
Fig.~\ref{fig:fsat_thr_results} shows the satellite fraction $f_{\rm sat}$ for the threshold samples as a function of the threshold mass $\log_{10}[M_{*,~{\rm lim}}/(h^{-2}M_\odot)]$ inferred by the $\Mprog$ and $\vpeak$ models \smrv{in the mini- and Shin-Uchuu simulations}.
For each threshold sample \smrv{and its} parameter sets \smrv{in the $1\sigma$ range}, $f_{\rm sat}$ is simply measured as
\begin{align}
    \fsat=\frac{\rm number~of~galaxies~hosted~by~satellite~subhalos}{\rm total~number~of~galaxies}.
\end{align}

The overall shape of the satellite fraction differs between the $\Mprog$ and $\vpeak$ models \smrv{in both simulations}.
The $\vpeak$ model yields a monotonically decreasing form with increasing stellar mass.
One would naively expect that the satellite fraction to be higher for lower mass galaxies, as the $\vpeak$ model predicts.
However, in the $\Mprog$ model, $\fsat$ as a function of the lower stellar mass limit has a single peak at \smrv{$\log_{10}[M_{*,~{\rm lim}}/(h^{-2}M_\odot)]=9.3\textrm{--}9.4$}.

The amplitude of galaxy clustering is strongly related to the satellite fraction.
For the threshold samples with $\log_{10}[M_{*,~{\rm lim}}/(h^{-2}M_\odot)]\geq10.4$, the ACFs of the $\Mprog$ and $\vpeak$ models \smrv{both agree with the observation well in the mini- and Shin-Uchuu simulations}.
The two models predict the satellite fractions very close to each other for these samples \smrv{in both simulations}.
For the threshold samples with \smrv{$9.2\leq\log_{10}[M_{*,~{\rm lim}}/(h^{-2}M_\odot)]\leq10.2$}, the $\Mprog$ model gives the higher $\fsat$ than the $\vpeak$ model as well as the amplitudes of the ACFs \smrv{in both simulations}.
This \smrv{means that} the higher amplitude of the correlation functions predicted by the $M_{\rm prog}$ model is due to the higher satellite fraction.

Fig.~\ref{fig:fsat_bin_results} is very similar to Fig.~\ref{fig:fsat_thr_results} but shows the inferred satellite fraction $\fsat$ for the stellar mass bin samples as a function of the mass range.
Compared to $\fsat$ for the threshold samples, the peak of the $\Mprog$ model is shifted to a  higher stellar mass, $\log_{10}[M_*/(h^{-2}M_\odot)]\simeq\smrv{9.4\textrm{--}10}$, because $\fsat$ for the threshold samples is the cumulation of $\fsat$ for the bin samples.

\cite{Knobel13} used the spectroscopic galaxy sample at $0.1<z<0.8$ from the zCOSMOS survey to evaluate the satellite fractions \citep[see][for the measurements at $z\simeq0$]{vdB08}.
They showed that the satellite fractions at $0.4<z<0.6$ and $0.6<z<0.8$ are not a monotonic function of the stellar mass, and have a peak at $\log_{10}[M_*/(h^{-2}M_\odot)]\simeq10.1$ and $10.3$, respectively.
This is \smrv{in the same trend as} the prediction of the $\Mprog$ model.

The stellar mass dependence of the satellite fraction of the $\Mprog$ model comes from \smrv{that of} $\zprog$.
The peak positions and shapes of the satellite fraction and $\zprog$ (Fig. \ref{fig:best_param_Mprog}) are very similar to each other.
This is because a larger $\zprog$ generally leads to a higher satellite fraction as discussed in Sec.~\ref{sec:xi_zprog}.

\section{Summary and Conclusion}
\label{sec:conclusion}
We have proposed a novel rank-ordering SHAM model using the progenitor virial mass of each subhalo at redshift $\zprog$, $\Mprog$, as a proxy of galaxy stellar mass at the time of observation.
In this model, the characteristic redshift $\zprog$ at which we evaluate $\Mprog$, and the scatter parameter $\sigma_M$ (see Eq.~\ref{eq:scatter_Mprog}) are the free fitting parameters.
The motivation of this model is related to the two-phase scenario of stellar mass growth in galaxies, i.e., in-situ star formation and ex-situ star accretion (see Sec.~\ref{sec:motivation}).

We have studied the $\zprog$-dependence of the 2PCFs \smrv{$\xi_M$} for the subhalo samples with the number density of $n_{\rm gal}=10^{-2}~h^3~{\rm Mpc}^{-3}$ at $z=0$ (Fig.~\ref{fig:xi_diff_z0p0}).
The $\zprog$-dependence can be understood by the variation of subhalo mass accretion histories and the subhalo mass stripping during accretion as shown in the halo occupation number (Fig.~\ref{fig:hod_Mzprog_z0p0}).
We have shown that the $\Mprog$ model with  certain $\zprog$ value gives the 2PCFs similar to or even more amplified than those by the $\vpeak$ model.
We expect the $\Mprog$ model to be able to well reproduce the observed galaxy clustering signal.

We have applied the $\Mprog$ and $\vpeak$ models \smrv{implemented in the mini- and Shin-Uchuu simulations} to the observed ACFs $\omega$ of the photo-$z$ selected galaxies at $z\simeq0.4$ from the Subaru HSC survey \smrv{(see Sec.~\ref{sec:obs})}.
Both models can reproduce the observed ACFs for the stellar mass threshold samples with $\log_{10}[M_{*,{\rm lim}}/(h^{-2}M_\odot)]\geq10.4$ (Fig. \ref{fig:w_results}).
We have also shown that the $\vpeak$ model underpredicts the amplitude of ACFs at $\lesssim1~\hiMpc$ and fails to match the observed ACFs for the samples of \smrv{$9.2\leq\log_{10}[M_{*,{\rm lim}}/(h^{-2}M_\odot)]\leq10.2$}.
On the other hand, the $\Mprog$ model gives higher amplitudes and \smrv{better} agrees with the observations for these samples.
\smrv{
The $\Mprog$ model in the high-resolution Shin-Uchuu simulation matches the observations down to $\simeq30~\hikpc$.
However, the predicted clustering amplitudes at $\lesssim30~\hikpc$ are lower than the observations.
The $\Mprog$ model in the lower resolution mini-Uchuu simulation underpredicts amplitudes at $\lesssim1~\hiMpc$ due to the resolution effect.
Therefore the model would need fine-tuning according to resolution.
It can be the inclusion of orphan galaxies, but its amount should be lower for the $\Mprog$ model than at least the $\vpeak$ model because the $\Mprog$ model can assign galaxies in more satellite subhalos.
}

We have found that $\zprog$ constrained by the observed ACFs has an interesting dependence on the stellar mass (Fig. \ref{fig:best_param_Mprog}).
The \smrv{obtained} $\zprog$ is lower toward the lowest and highest stellar mass ranges and has a single peak at $\log_{10}[M_*/(h^{-2}M_\odot)]\simeq9.9$.
This trend is qualitatively consistent with the in-/ex-situ scenario of stellar mass growth.
It is clearly important to quantitatively examine whether the obtained values of $\zprog$ are physically reasonable.
Specifically, investigating what events related to stellar mass growth happened at $z=\zprog$ should be an interesting topic.
The recent cosmological simulations of galaxy formation 
(e.g., EAGLE; \citealt{Crain15, Schaye15}; Illustris TNG; \citealt{Weinberger17, Pillepich18_tng}; FIREbox; \citealt{Feldmann23})
might give clues to understanding the physical origin of the \smrv{obtained} $\zprog$ values, although it is beyond the scope of this work.

We have studied the inferred satellite fractions in Figs.~\ref{fig:fsat_thr_results} and  \ref{fig:fsat_bin_results}.
The successful agreement of the $\Mprog$ model with the observed ACFs for the samples with the thresholds of \smrv{$9.2\leq\log_{10}[M_{*,{\rm lim}}/(h^{-2}M_\odot)]\leq10.2$} is attributed
to the higher satellite fraction than in the $\vpeak$ model.
For the other mass-threshold samples for which both models can reproduce the observed ACFs, the predicted satellite fractions from the two models agree with each other.
Thus the satellite fraction is a crucial factor for determining the strength of galaxy clustering.

In future work, we plan to examine the $\Mprog$ SHAM model by comparing it with observed galaxy statistics as a function of the stellar mass including clustering measurements at various redshifts \citep[e.g.,][]{Yang12,Ishikawa20,Shuntov22}, 
the satellite fractions as a function of the stellar mass \citep{vdB08,Knobel13}, 
the mass profiles around galaxies \citep{mandelbaum06,Leauthaud12} and 
the group statistics \citep{Hearin13b}.

With only one more free parameter than the $\vpeak$ model, the  $\Mprog$ model is shown to be highly flexible and can more faithfully reproduce the observed ACFs, providing a more physical way to interpret the observed clustering measurements.
The findings in this paper would be an important step toward accurate modeling of the galaxy-halo connection.

\section*{Acknowledgements}
\smrvv{We would like to appreciate the anonymous referee for constructive and valuable comments that helped us improve this paper.}
We would like to thank Yoshiki Matsuoka for useful discussions in the early stage of this work, Tomoaki Ishiyama for the details of the Uchuu simulations, and Peter Behroozi and Masato Shirasaki for useful comments on the manuscript.
The calculations in part were carried out on Cray XC50 at Center for Computational Astrophysics, National Astronomical Observatory of Japan.
This work was supported in part by JSPS KAKENHI Grant Numbers JP19H00677, JP21H05465, JP22K03644 (SM), JP21K13956 (DK), and JP23K13145 (SI). YTL acknowledges support from the National Science and Technology Council of Taiwan under grants MOST 111-2112-M-001-043 and MOST 110-2112-M-001-004.

\section*{Data Availability}
The Uchuu simulations are publicly available\footnote{\url{http://skiesanduniverses.org/Simulations/Uchuu/}}.
\smrvv{The ACFs data of the HSC galaxies are available on the first author's web page\footnote{\url{https://sites.google.com/view/smasaki}}.}
Other data presented in this paper can be provided by the authors upon request.



\bibliographystyle{mnras}
\bibliography{lssref} 

\begin{thebibliography}{}
\makeatletter
\relax
\def\mn@urlcharsother{\let\do\@makeother \do\$\do\&\do\#\do\^\do\_\do\%\do\~}
\def\mn@doi{\begingroup\mn@urlcharsother \@ifnextchar [ {\mn@doi@}
  {\mn@doi@[]}}
\def\mn@doi@[#1]#2{\def\@tempa{#1}\ifx\@tempa\@empty \href
  {http://dx.doi.org/#2} {doi:#2}\else \href {http://dx.doi.org/#2} {#1}\fi
  \endgroup}
\def\mn@eprint#1#2{\mn@eprint@#1:#2::\@nil}
\def\mn@eprint@arXiv#1{\href {http://arxiv.org/abs/#1} {{\tt arXiv:#1}}}
\def\mn@eprint@dblp#1{\href {http://dblp.uni-trier.de/rec/bibtex/#1.xml}
  {dblp:#1}}
\def\mn@eprint@#1:#2:#3:#4\@nil{\def\@tempa {#1}\def\@tempb {#2}\def\@tempc
  {#3}\ifx \@tempc \@empty \let \@tempc \@tempb \let \@tempb \@tempa \fi \ifx
  \@tempb \@empty \def\@tempb {arXiv}\fi \@ifundefined
  {mn@eprint@\@tempb}{\@tempb:\@tempc}{\expandafter \expandafter \csname
  mn@eprint@\@tempb\endcsname \expandafter{\@tempc}}}

\bibitem[\protect\citeauthoryear{{Ahn} et~al.,}{{Ahn} et~al.}{2014}]{Ahn14}
{Ahn} C.~P.,  et~al., 2014, \mn@doi [\apjs] {10.1088/0067-0049/211/2/17}, \href
  {https://ui.adsabs.harvard.edu/abs/2014ApJS..211...17A} {211, 17}

\bibitem[\protect\citeauthoryear{{Aihara} et~al.,}{{Aihara}
  et~al.}{2018}]{Aihara18}
{Aihara} H.,  et~al., 2018, \mn@doi [\pasj] {10.1093/pasj/psx066}, \href
  {https://ui.adsabs.harvard.edu/abs/2018PASJ...70S...4A} {70, S4}

\bibitem[\protect\citeauthoryear{{Alam}, {Miyatake}, {More}, {Ho}  \&
  {Mandelbaum}}{{Alam} et~al.}{2017}]{Alam17}
{Alam} S.,  {Miyatake} H.,  {More} S.,  {Ho} S.,   {Mandelbaum} R.,  2017,
  \mn@doi [\mnras] {10.1093/mnras/stw3056}, \href
  {https://ui.adsabs.harvard.edu/abs/2017MNRAS.465.4853A} {465, 4853}

\bibitem[\protect\citeauthoryear{{Amodeo} et~al.,}{{Amodeo}
  et~al.}{2021}]{Amodeo21}
{Amodeo} S.,  et~al., 2021, \mn@doi [\prd] {10.1103/PhysRevD.103.063514}, \href
  {https://ui.adsabs.harvard.edu/abs/2021PhRvD.103f3514A} {103, 063514}

\bibitem[\protect\citeauthoryear{{Amon} et~al.,}{{Amon} et~al.}{2023}]{Amon23}
{Amon} A.,  et~al., 2023, \mn@doi [\mnras] {10.1093/mnras/stac2938}, \href
  {https://ui.adsabs.harvard.edu/abs/2023MNRAS.518..477A} {518, 477}

\bibitem[\protect\citeauthoryear{{Behroozi}, {Wechsler}  \& {Wu}}{{Behroozi}
  et~al.}{2013a}]{Behroozi:2013}
{Behroozi} P.~S.,  {Wechsler} R.~H.,   {Wu} H.-Y.,  2013a, \mn@doi [\apj]
  {10.1088/0004-637X/762/2/109}, \href
  {http://adsabs.harvard.edu/abs/2013ApJ...762..109B} {762, 109}

\bibitem[\protect\citeauthoryear{{Behroozi}, {Wechsler}, {Wu}, {Busha},
  {Klypin}  \& {Primack}}{{Behroozi} et~al.}{2013b}]{Behroozi13_consistenttree}
{Behroozi} P.~S.,  {Wechsler} R.~H.,  {Wu} H.-Y.,  {Busha} M.~T.,  {Klypin}
  A.~A.,   {Primack} J.~R.,  2013b, \mn@doi [\apj]
  {10.1088/0004-637X/763/1/18}, \href
  {https://ui.adsabs.harvard.edu/abs/2013ApJ...763...18B} {763, 18}

\bibitem[\protect\citeauthoryear{{Behroozi}, {Wechsler}  \&
  {Conroy}}{{Behroozi} et~al.}{2013c}]{Behroozi13_wc}
{Behroozi} P.~S.,  {Wechsler} R.~H.,   {Conroy} C.,  2013c, \mn@doi [\apj]
  {10.1088/0004-637X/770/1/57}, \href
  {https://ui.adsabs.harvard.edu/abs/2013ApJ...770...57B} {770, 57}

\bibitem[\protect\citeauthoryear{{Behroozi}, {Wechsler}, {Lu}, {Hahn}, {Busha},
  {Klypin}  \& {Primack}}{{Behroozi} et~al.}{2014}]{Behroozi14}
{Behroozi} P.~S.,  {Wechsler} R.~H.,  {Lu} Y.,  {Hahn} O.,  {Busha} M.~T.,
  {Klypin} A.,   {Primack} J.~R.,  2014, \mn@doi [\apj]
  {10.1088/0004-637X/787/2/156}, \href
  {https://ui.adsabs.harvard.edu/abs/2014ApJ...787..156B} {787, 156}

\bibitem[\protect\citeauthoryear{{Behroozi}, {Wechsler}, {Hearin}  \&
  {Conroy}}{{Behroozi} et~al.}{2019}]{Behroozi19}
{Behroozi} P.,  {Wechsler} R.~H.,  {Hearin} A.~P.,   {Conroy} C.,  2019,
  \mn@doi [\mnras] {10.1093/mnras/stz1182}, \href
  {https://ui.adsabs.harvard.edu/abs/2019MNRAS.488.3143B} {488, 3143}

\bibitem[\protect\citeauthoryear{{Bell} et~al.,}{{Bell} et~al.}{2005}]{Bell05}
{Bell} E.~F.,  et~al., 2005, \mn@doi [\apj] {10.1086/429552}, \href
  {https://ui.adsabs.harvard.edu/abs/2005ApJ...625...23B} {625, 23}

\bibitem[\protect\citeauthoryear{{Brinchmann} \& {Ellis}}{{Brinchmann} \&
  {Ellis}}{2000}]{Brinchmann00}
{Brinchmann} J.,  {Ellis} R.~S.,  2000, \mn@doi [\apjl] {10.1086/312738}, \href
  {https://ui.adsabs.harvard.edu/abs/2000ApJ...536L..77B} {536, L77}

\bibitem[\protect\citeauthoryear{{Bundy} et~al.,}{{Bundy}
  et~al.}{2006}]{Bundy06}
{Bundy} K.,  et~al., 2006, \mn@doi [\apj] {10.1086/507456}, \href
  {https://ui.adsabs.harvard.edu/abs/2006ApJ...651..120B} {651, 120}

\bibitem[\protect\citeauthoryear{{Campbell}, {van den Bosch}, {Padmanabhan},
  {Mao}, {Zentner}, {Lange}, {Jiang}  \& {Villarreal}}{{Campbell}
  et~al.}{2018}]{Campbell18}
{Campbell} D.,  {van den Bosch} F.~C.,  {Padmanabhan} N.,  {Mao} Y.-Y.,
  {Zentner} A.~R.,  {Lange} J.~U.,  {Jiang} F.,   {Villarreal} A.~S.,  2018,
  \mn@doi [\mnras] {10.1093/mnras/sty495}, \href
  {https://ui.adsabs.harvard.edu/abs/2018MNRAS.477..359C} {477, 359}

\bibitem[\protect\citeauthoryear{{Cannarozzo} et~al.,}{{Cannarozzo}
  et~al.}{2023}]{Cannarozzo23}
{Cannarozzo} C.,  et~al., 2023, \mn@doi [\mnras] {10.1093/mnras/stac3023},
  \href {https://ui.adsabs.harvard.edu/abs/2023MNRAS.520.5651C} {520, 5651}

\bibitem[\protect\citeauthoryear{{Chaves-Montero}, {Angulo}, {Schaye},
  {Schaller}, {Crain}, {Furlong}  \& {Theuns}}{{Chaves-Montero}
  et~al.}{2016}]{Chaves-Montero16}
{Chaves-Montero} J.,  {Angulo} R.~E.,  {Schaye} J.,  {Schaller} M.,  {Crain}
  R.~A.,  {Furlong} M.,   {Theuns} T.,  2016, \mn@doi [\mnras]
  {10.1093/mnras/stw1225}, \href
  {https://ui.adsabs.harvard.edu/abs/2016MNRAS.460.3100C} {460, 3100}

\bibitem[\protect\citeauthoryear{{Chuang} \& {Lin}}{{Chuang} \&
  {Lin}}{2023}]{Chuang23}
{Chuang} C.-Y.,  {Lin} Y.-T.,  2023, \mn@doi [\apj] {10.3847/1538-4357/acb5f3},
  \href {https://ui.adsabs.harvard.edu/abs/2023ApJ...944..207C} {944, 207}

\bibitem[\protect\citeauthoryear{{Conroy}, {Wechsler}  \& {Kravtsov}}{{Conroy}
  et~al.}{2006}]{conroy06}
{Conroy} C.,  {Wechsler} R.~H.,   {Kravtsov} A.~V.,  2006, \mn@doi [\apj]
  {10.1086/503602}, \href {http://adsabs.harvard.edu/abs/2006ApJ...647..201C}
  {647, 201}

\bibitem[\protect\citeauthoryear{{Contreras}, {Chaves-Montero}  \&
  {Angulo}}{{Contreras} et~al.}{2023}]{Contreras23}
{Contreras} S.,  {Chaves-Montero} J.,   {Angulo} R.~E.,  2023, \mn@doi [arXiv
  e-prints] {10.48550/arXiv.2305.09637}, \href
  {https://ui.adsabs.harvard.edu/abs/2023arXiv230509637C} {p. arXiv:2305.09637}

\bibitem[\protect\citeauthoryear{{Cooray} \& {Sheth}}{{Cooray} \&
  {Sheth}}{2002}]{Cooray02}
{Cooray} A.,  {Sheth} R.,  2002, \mn@doi [\physrep]
  {10.1016/S0370-1573(02)00276-4}, \href
  {http://adsabs.harvard.edu/abs/2002PhR...372....1C} {372, 1}

\bibitem[\protect\citeauthoryear{{Cowie}, {Songaila}, {Hu}  \& {Cohen}}{{Cowie}
  et~al.}{1996}]{Cowie96}
{Cowie} L.~L.,  {Songaila} A.,  {Hu} E.~M.,   {Cohen} J.~G.,  1996, \mn@doi
  [\aj] {10.1086/118058}, \href
  {https://ui.adsabs.harvard.edu/abs/1996AJ....112..839C} {112, 839}

\bibitem[\protect\citeauthoryear{{Crain} et~al.,}{{Crain}
  et~al.}{2015}]{Crain15}
{Crain} R.~A.,  et~al., 2015, \mn@doi [\mnras] {10.1093/mnras/stv725}, \href
  {https://ui.adsabs.harvard.edu/abs/2015MNRAS.450.1937C} {450, 1937}

\bibitem[\protect\citeauthoryear{{Davison}, {Norris}, {Pfeffer}, {Davies}  \&
  {Crain}}{{Davison} et~al.}{2020}]{Davison20}
{Davison} T.~A.,  {Norris} M.~A.,  {Pfeffer} J.~L.,  {Davies} J.~J.,   {Crain}
  R.~A.,  2020, \mn@doi [\mnras] {10.1093/mnras/staa1816}, \href
  {https://ui.adsabs.harvard.edu/abs/2020MNRAS.497...81D} {497, 81}

\bibitem[\protect\citeauthoryear{{Dong-P{\'a}ez} et~al.,}{{Dong-P{\'a}ez}
  et~al.}{2022}]{DongPaez22}
{Dong-P{\'a}ez} C.~A.,  et~al., 2022, arXiv e-prints, \href
  {https://ui.adsabs.harvard.edu/abs/2022arXiv220800540D} {p. arXiv:2208.00540}

\bibitem[\protect\citeauthoryear{{Feldmann} et~al.,}{{Feldmann}
  et~al.}{2023}]{Feldmann23}
{Feldmann} R.,  et~al., 2023, \mn@doi [\mnras] {10.1093/mnras/stad1205}, \href
  {https://ui.adsabs.harvard.edu/abs/2023MNRAS.522.3831F} {522, 3831}

\bibitem[\protect\citeauthoryear{{Gao} \& {White}}{{Gao} \&
  {White}}{2007}]{Gao07}
{Gao} L.,  {White} S. D.~M.,  2007, \mn@doi [\mnras]
  {10.1111/j.1745-3933.2007.00292.x}, \href
  {https://ui.adsabs.harvard.edu/abs/2007MNRAS.377L...5G} {377, L5}

\bibitem[\protect\citeauthoryear{{Guo} \& {White}}{{Guo} \&
  {White}}{2014}]{Guo14}
{Guo} Q.,  {White} S.,  2014, \mn@doi [\mnras] {10.1093/mnras/stt2116}, \href
  {https://ui.adsabs.harvard.edu/abs/2014MNRAS.437.3228G} {437, 3228}

\bibitem[\protect\citeauthoryear{{Guzm{\'a}n}, {Gallego}, {Koo}, {Phillips},
  {Lowenthal}, {Faber}, {Illingworth}  \& {Vogt}}{{Guzm{\'a}n}
  et~al.}{1997}]{Guzman97}
{Guzm{\'a}n} R.,  {Gallego} J.,  {Koo} D.~C.,  {Phillips} A.~C.,  {Lowenthal}
  J.~D.,  {Faber} S.~M.,  {Illingworth} G.~D.,   {Vogt} N.~P.,  1997, \mn@doi
  [\apj] {10.1086/304797}, \href
  {https://ui.adsabs.harvard.edu/abs/1997ApJ...489..559G} {489, 559}

\bibitem[\protect\citeauthoryear{{Hand}, {Feng}, {Beutler}, {Li}, {Modi},
  {Seljak}  \& {Slepian}}{{Hand} et~al.}{2018}]{Hand18}
{Hand} N.,  {Feng} Y.,  {Beutler} F.,  {Li} Y.,  {Modi} C.,  {Seljak} U.,
  {Slepian} Z.,  2018, \mn@doi [\aj] {10.3847/1538-3881/aadae0}, \href
  {https://ui.adsabs.harvard.edu/abs/2018AJ....156..160H} {156, 160}

\bibitem[\protect\citeauthoryear{{Hartlap}, {Simon}  \& {Schneider}}{{Hartlap}
  et~al.}{2007}]{Hartlap07}
{Hartlap} J.,  {Simon} P.,   {Schneider} P.,  2007, \mn@doi [\aap]
  {10.1051/0004-6361:20066170}, \href
  {https://ui.adsabs.harvard.edu/abs/2007A&A...464..399H} {464, 399}

\bibitem[\protect\citeauthoryear{{Hearin} \& {Watson}}{{Hearin} \&
  {Watson}}{2013}]{Hearin13}
{Hearin} A.~P.,  {Watson} D.~F.,  2013, \mn@doi [\mnras]
  {10.1093/mnras/stt1374}, \href
  {https://ui.adsabs.harvard.edu/abs/2013MNRAS.435.1313H} {435, 1313}

\bibitem[\protect\citeauthoryear{{Hearin}, {Zentner}, {Berlind}  \&
  {Newman}}{{Hearin} et~al.}{2013}]{Hearin13b}
{Hearin} A.~P.,  {Zentner} A.~R.,  {Berlind} A.~A.,   {Newman} J.~A.,  2013,
  \mn@doi [\mnras] {10.1093/mnras/stt755}, \href
  {https://ui.adsabs.harvard.edu/abs/2013MNRAS.433..659H} {433, 659}

\bibitem[\protect\citeauthoryear{{Ishikawa} et~al.,}{{Ishikawa}
  et~al.}{2020}]{Ishikawa20}
{Ishikawa} S.,  et~al., 2020, \mn@doi [\apj] {10.3847/1538-4357/abbd95}, \href
  {https://ui.adsabs.harvard.edu/abs/2020ApJ...904..128I} {904, 128}

\bibitem[\protect\citeauthoryear{{Ishiyama}, {Fukushige}  \&
  {Makino}}{{Ishiyama} et~al.}{2009}]{ishiyama09}
{Ishiyama} T.,  {Fukushige} T.,   {Makino} J.,  2009, \mn@doi [\pasj]
  {10.1093/pasj/61.6.1319}, \href
  {https://ui.adsabs.harvard.edu/abs/2009PASJ...61.1319I} {61, 1319}

\bibitem[\protect\citeauthoryear{{Ishiyama} et~al.,}{{Ishiyama}
  et~al.}{2021}]{Ishiyama21}
{Ishiyama} T.,  et~al., 2021, \mn@doi [\mnras] {10.1093/mnras/stab1755}, \href
  {https://ui.adsabs.harvard.edu/abs/2021MNRAS.506.4210I} {506, 4210}

\bibitem[\protect\citeauthoryear{{Jimenez}, {Panter}, {Heavens}  \&
  {Verde}}{{Jimenez} et~al.}{2005}]{Jimenez05}
{Jimenez} R.,  {Panter} B.,  {Heavens} A.~F.,   {Verde} L.,  2005, \mn@doi
  [\mnras] {10.1111/j.1365-2966.2004.08469.x}, \href
  {https://ui.adsabs.harvard.edu/abs/2005MNRAS.356..495J} {356, 495}

\bibitem[\protect\citeauthoryear{{Juneau} et~al.,}{{Juneau}
  et~al.}{2005}]{Juneau05}
{Juneau} S.,  et~al., 2005, \mn@doi [\apjl] {10.1086/427937}, \href
  {https://ui.adsabs.harvard.edu/abs/2005ApJ...619L.135J} {619, L135}

\bibitem[\protect\citeauthoryear{{Knobel} et~al.,}{{Knobel}
  et~al.}{2013}]{Knobel13}
{Knobel} C.,  et~al., 2013, \mn@doi [\apj] {10.1088/0004-637X/769/1/24}, \href
  {https://ui.adsabs.harvard.edu/abs/2013ApJ...769...24K} {769, 24}

\bibitem[\protect\citeauthoryear{{Kodama} et~al.,}{{Kodama}
  et~al.}{2004}]{Kodama04}
{Kodama} T.,  et~al., 2004, \mn@doi [\mnras]
  {10.1111/j.1365-2966.2004.07711.x}, \href
  {https://ui.adsabs.harvard.edu/abs/2004MNRAS.350.1005K} {350, 1005}

\bibitem[\protect\citeauthoryear{{Kravtsov}, {Berlind}, {Wechsler}, {Klypin},
  {Gottl{\"o}ber}, {Allgood}  \& {Primack}}{{Kravtsov}
  et~al.}{2004}]{Kravtsov04}
{Kravtsov} A.~V.,  {Berlind} A.~A.,  {Wechsler} R.~H.,  {Klypin} A.~A.,
  {Gottl{\"o}ber} S.,  {Allgood} B.,   {Primack} J.~R.,  2004, \mn@doi [\apj]
  {10.1086/420959}, \href
  {https://ui.adsabs.harvard.edu/abs/2004ApJ...609...35K} {609, 35}

\bibitem[\protect\citeauthoryear{{Lackner}, {Cen}, {Ostriker}  \&
  {Joung}}{{Lackner} et~al.}{2012}]{Lackner12}
{Lackner} C.~N.,  {Cen} R.,  {Ostriker} J.~P.,   {Joung} M.~R.,  2012, \mn@doi
  [\mnras] {10.1111/j.1365-2966.2012.21525.x}, \href
  {https://ui.adsabs.harvard.edu/abs/2012MNRAS.425..641L} {425, 641}

\bibitem[\protect\citeauthoryear{{Landy} \& {Szalay}}{{Landy} \&
  {Szalay}}{1993}]{LS93}
{Landy} S.~D.,  {Szalay} A.~S.,  1993, \mn@doi [\apj] {10.1086/172900}, \href
  {https://ui.adsabs.harvard.edu/abs/1993ApJ...412...64L} {412, 64}

\bibitem[\protect\citeauthoryear{{Lange}, {Yang}, {Guo}, {Luo}  \& {van den
  Bosch}}{{Lange} et~al.}{2019}]{Lange19}
{Lange} J.~U.,  {Yang} X.,  {Guo} H.,  {Luo} W.,   {van den Bosch} F.~C.,
  2019, \mn@doi [\mnras] {10.1093/mnras/stz2124}, \href
  {https://ui.adsabs.harvard.edu/abs/2019MNRAS.488.5771L} {488, 5771}

\bibitem[\protect\citeauthoryear{{Lange}, {Leauthaud}, {Singh}, {Guo}, {Zhou},
  {Smith}  \& {Cyr-Racine}}{{Lange} et~al.}{2021}]{Lange21}
{Lange} J.~U.,  {Leauthaud} A.,  {Singh} S.,  {Guo} H.,  {Zhou} R.,  {Smith}
  T.~L.,   {Cyr-Racine} F.-Y.,  2021, \mn@doi [\mnras] {10.1093/mnras/stab189},
  \href {https://ui.adsabs.harvard.edu/abs/2021MNRAS.502.2074L} {502, 2074}

\bibitem[\protect\citeauthoryear{{Leauthaud} et~al.,}{{Leauthaud}
  et~al.}{2012}]{Leauthaud12}
{Leauthaud} A.,  et~al., 2012, \mn@doi [\apj] {10.1088/0004-637X/744/2/159},
  \href {https://ui.adsabs.harvard.edu/abs/2012ApJ...744..159L} {744, 159}

\bibitem[\protect\citeauthoryear{{Leauthaud} et~al.,}{{Leauthaud}
  et~al.}{2017}]{Leauthaud17}
{Leauthaud} A.,  et~al., 2017, \mn@doi [\mnras] {10.1093/mnras/stx258}, \href
  {https://ui.adsabs.harvard.edu/abs/2017MNRAS.467.3024L} {467, 3024}

\bibitem[\protect\citeauthoryear{{Lehmann}, {Mao}, {Becker}, {Skillman}  \&
  {Wechsler}}{{Lehmann} et~al.}{2017}]{Lehmann17}
{Lehmann} B.~V.,  {Mao} Y.-Y.,  {Becker} M.~R.,  {Skillman} S.~W.,   {Wechsler}
  R.~H.,  2017, \mn@doi [\apj] {10.3847/1538-4357/834/1/37}, \href
  {https://ui.adsabs.harvard.edu/abs/2017ApJ...834...37L} {834, 37}

\bibitem[\protect\citeauthoryear{{Lesgourgues}}{{Lesgourgues}}{2011}]{class1}
{Lesgourgues} J.,  2011, preprint, \href
  {http://adsabs.harvard.edu/abs/2011arXiv1104.2932L} {} (\mn@eprint {arXiv}
  {1104.2932})

\bibitem[\protect\citeauthoryear{{Limber}}{{Limber}}{1953}]{Limber53}
{Limber} D.~N.,  1953, \mn@doi [\apj] {10.1086/145672}, \href
  {https://ui.adsabs.harvard.edu/abs/1953ApJ...117..134L} {117, 134}

\bibitem[\protect\citeauthoryear{{Mandelbaum}, {Seljak}, {Kauffmann}, {Hirata}
  \& {Brinkmann}}{{Mandelbaum} et~al.}{2006}]{mandelbaum06}
{Mandelbaum} R.,  {Seljak} U.,  {Kauffmann} G.,  {Hirata} C.~M.,   {Brinkmann}
  J.,  2006, \mn@doi [MNRAS] {10.1111/j.1365-2966.2006.10156.x}, \href
  {http://adsabs.harvard.edu/abs/2006MNRAS.368..715M} {368, 715}

\bibitem[\protect\citeauthoryear{{Mansfield} \& {Avestruz}}{{Mansfield} \&
  {Avestruz}}{2021}]{Masfield21}
{Mansfield} P.,  {Avestruz} C.,  2021, \mn@doi [\mnras]
  {10.1093/mnras/staa3388}, \href
  {https://ui.adsabs.harvard.edu/abs/2021MNRAS.500.3309M} {500, 3309}

\bibitem[\protect\citeauthoryear{{Masaki}, {Hikage}, {Takada}, {Spergel}  \&
  {Sugiyama}}{{Masaki} et~al.}{2013a}]{masaki13}
{Masaki} S.,  {Hikage} C.,  {Takada} M.,  {Spergel} D.~N.,   {Sugiyama} N.,
  2013a, \mn@doi [MNRAS] {10.1093/mnras/stt981}, \href
  {http://adsabs.harvard.edu/abs/2013MNRAS.433.3506M} {433, 3506}

\bibitem[\protect\citeauthoryear{{Masaki}, {Lin}  \& {Yoshida}}{{Masaki}
  et~al.}{2013b}]{Masaki13b}
{Masaki} S.,  {Lin} Y.-T.,   {Yoshida} N.,  2013b, \mn@doi [\mnras]
  {10.1093/mnras/stt1729}, \href
  {https://ui.adsabs.harvard.edu/abs/2013MNRAS.436.2286M} {436, 2286}

\bibitem[\protect\citeauthoryear{{McBride}, {Fakhouri}  \& {Ma}}{{McBride}
  et~al.}{2009}]{McBride09}
{McBride} J.,  {Fakhouri} O.,   {Ma} C.-P.,  2009, \mn@doi [\mnras]
  {10.1111/j.1365-2966.2009.15329.x}, \href
  {https://ui.adsabs.harvard.edu/abs/2009MNRAS.398.1858M} {398, 1858}

\bibitem[\protect\citeauthoryear{{Moster}, {Naab}  \& {White}}{{Moster}
  et~al.}{2013}]{Moster13}
{Moster} B.~P.,  {Naab} T.,   {White} S. D.~M.,  2013, \mn@doi [\mnras]
  {10.1093/mnras/sts261}, \href
  {https://ui.adsabs.harvard.edu/abs/2013MNRAS.428.3121M} {428, 3121}

\bibitem[\protect\citeauthoryear{{Moster}, {Naab}  \& {White}}{{Moster}
  et~al.}{2018}]{Moster18}
{Moster} B.~P.,  {Naab} T.,   {White} S. D.~M.,  2018, \mn@doi [\mnras]
  {10.1093/mnras/sty655}, \href
  {https://ui.adsabs.harvard.edu/abs/2018MNRAS.477.1822M} {477, 1822}

\bibitem[\protect\citeauthoryear{{Neistein}, {van den Bosch}  \&
  {Dekel}}{{Neistein} et~al.}{2006}]{Neistein06}
{Neistein} E.,  {van den Bosch} F.~C.,   {Dekel} A.,  2006, \mn@doi [\mnras]
  {10.1111/j.1365-2966.2006.10918.x}, \href
  {https://ui.adsabs.harvard.edu/abs/2006MNRAS.372..933N} {372, 933}

\bibitem[\protect\citeauthoryear{{Norberg}, {Baugh}, {Gazta{\~n}aga}  \&
  {Croton}}{{Norberg} et~al.}{2009}]{Norberg09}
{Norberg} P.,  {Baugh} C.~M.,  {Gazta{\~n}aga} E.,   {Croton} D.~J.,  2009,
  \mn@doi [\mnras] {10.1111/j.1365-2966.2009.14389.x}, \href
  {https://ui.adsabs.harvard.edu/abs/2009MNRAS.396...19N} {396, 19}

\bibitem[\protect\citeauthoryear{{Nuza} et~al.,}{{Nuza} et~al.}{2013}]{Nuza13}
{Nuza} S.~E.,  et~al., 2013, \mn@doi [\mnras] {10.1093/mnras/stt513}, \href
  {https://ui.adsabs.harvard.edu/abs/2013MNRAS.432..743N} {432, 743}

\bibitem[\protect\citeauthoryear{{Okumura}, {Hayashi}, {Chiu}, {Lin}, {Osato},
  {Hsieh}  \& {Lin}}{{Okumura} et~al.}{2021}]{Okumura21}
{Okumura} T.,  {Hayashi} M.,  {Chiu} I.~N.,  {Lin} Y.-T.,  {Osato} K.,  {Hsieh}
  B.-C.,   {Lin} S.-C.,  2021, \mn@doi [\pasj] {10.1093/pasj/psab068}, \href
  {https://ui.adsabs.harvard.edu/abs/2021PASJ...73.1186O} {73, 1186}

\bibitem[\protect\citeauthoryear{{Oser}, {Ostriker}, {Naab}, {Johansson}  \&
  {Burkert}}{{Oser} et~al.}{2010}]{Oser10}
{Oser} L.,  {Ostriker} J.~P.,  {Naab} T.,  {Johansson} P.~H.,   {Burkert} A.,
  2010, \mn@doi [\apj] {10.1088/0004-637X/725/2/2312}, \href
  {https://ui.adsabs.harvard.edu/abs/2010ApJ...725.2312O} {725, 2312}

\bibitem[\protect\citeauthoryear{{Peebles} \& {Groth}}{{Peebles} \&
  {Groth}}{1976}]{Peebles76}
{Peebles} P.~J.~E.,  {Groth} E.~J.,  1976, \aap, \href
  {https://ui.adsabs.harvard.edu/abs/1976A&A....53..131P} {53, 131}

\bibitem[\protect\citeauthoryear{{Pillepich}, {Madau}  \& {Mayer}}{{Pillepich}
  et~al.}{2015}]{Pillepich15}
{Pillepich} A.,  {Madau} P.,   {Mayer} L.,  2015, \mn@doi [\apj]
  {10.1088/0004-637X/799/2/184}, \href
  {https://ui.adsabs.harvard.edu/abs/2015ApJ...799..184P} {799, 184}

\bibitem[\protect\citeauthoryear{{Pillepich} et~al.,}{{Pillepich}
  et~al.}{2018a}]{Pillepich18_tng}
{Pillepich} A.,  et~al., 2018a, \mn@doi [\mnras] {10.1093/mnras/stx2656}, \href
  {https://ui.adsabs.harvard.edu/abs/2018MNRAS.473.4077P} {473, 4077}

\bibitem[\protect\citeauthoryear{{Pillepich} et~al.,}{{Pillepich}
  et~al.}{2018b}]{Pillepich18}
{Pillepich} A.,  et~al., 2018b, \mn@doi [\mnras] {10.1093/mnras/stx3112}, \href
  {https://ui.adsabs.harvard.edu/abs/2018MNRAS.475..648P} {475, 648}

\bibitem[\protect\citeauthoryear{{Planck Collaboration} et~al.,}{{Planck
  Collaboration} et~al.}{2020}]{Planck18}
{Planck Collaboration} et~al., 2020, \mn@doi [\aap]
  {10.1051/0004-6361/201833910}, \href
  {https://ui.adsabs.harvard.edu/abs/2020A&A...641A...6P} {641, A6}

\bibitem[\protect\citeauthoryear{{Reddick}, {Wechsler}, {Tinker}  \&
  {Behroozi}}{{Reddick} et~al.}{2013}]{Reddick13}
{Reddick} R.~M.,  {Wechsler} R.~H.,  {Tinker} J.~L.,   {Behroozi} P.~S.,  2013,
  \mn@doi [\apj] {10.1088/0004-637X/771/1/30}, \href
  {https://ui.adsabs.harvard.edu/abs/2013ApJ...771...30R} {771, 30}

\bibitem[\protect\citeauthoryear{{Reid}, {Seo}, {Leauthaud}, {Tinker}  \&
  {White}}{{Reid} et~al.}{2014}]{Reid14}
{Reid} B.~A.,  {Seo} H.-J.,  {Leauthaud} A.,  {Tinker} J.~L.,   {White} M.,
  2014, \mn@doi [\mnras] {10.1093/mnras/stu1391}, \href
  {https://ui.adsabs.harvard.edu/abs/2014MNRAS.444..476R} {444, 476}

\bibitem[\protect\citeauthoryear{{Roche}, {Eales}, {Hippelein}  \&
  {Willott}}{{Roche} et~al.}{1999}]{Roche99}
{Roche} N.,  {Eales} S.~A.,  {Hippelein} H.,   {Willott} C.~J.,  1999, \mn@doi
  [\mnras] {10.1046/j.1365-8711.1999.02536.x}, \href
  {https://ui.adsabs.harvard.edu/abs/1999MNRAS.306..538R} {306, 538}

\bibitem[\protect\citeauthoryear{{Rodriguez-Gomez} et~al.,}{{Rodriguez-Gomez}
  et~al.}{2016}]{RG16}
{Rodriguez-Gomez} V.,  et~al., 2016, \mn@doi [\mnras] {10.1093/mnras/stw456},
  \href {https://ui.adsabs.harvard.edu/abs/2016MNRAS.458.2371R} {458, 2371}

\bibitem[\protect\citeauthoryear{{Rodr{\'\i}guez-Torres}
  et~al.,}{{Rodr{\'\i}guez-Torres} et~al.}{2016}]{RT16}
{Rodr{\'\i}guez-Torres} S.~A.,  et~al., 2016, \mn@doi [\mnras]
  {10.1093/mnras/stw1014}, \href
  {https://ui.adsabs.harvard.edu/abs/2016MNRAS.460.1173R} {460, 1173}

\bibitem[\protect\citeauthoryear{{Saito} et~al.,}{{Saito}
  et~al.}{2016}]{Saito16}
{Saito} S.,  et~al., 2016, \mn@doi [\mnras] {10.1093/mnras/stw1080}, \href
  {https://ui.adsabs.harvard.edu/abs/2016MNRAS.460.1457S} {460, 1457}

\bibitem[\protect\citeauthoryear{{Schaye} et~al.,}{{Schaye}
  et~al.}{2015}]{Schaye15}
{Schaye} J.,  et~al., 2015, \mn@doi [\mnras] {10.1093/mnras/stu2058}, \href
  {https://ui.adsabs.harvard.edu/abs/2015MNRAS.446..521S} {446, 521}

\bibitem[\protect\citeauthoryear{{Shuntov} et~al.,}{{Shuntov}
  et~al.}{2022}]{Shuntov22}
{Shuntov} M.,  et~al., 2022, \mn@doi [\aap] {10.1051/0004-6361/202243136},
  \href {https://ui.adsabs.harvard.edu/abs/2022A&A...664A..61S} {664, A61}

\bibitem[\protect\citeauthoryear{{Simon}}{{Simon}}{2007}]{Simon07}
{Simon} P.,  2007, \mn@doi [\aap] {10.1051/0004-6361:20066352}, \href
  {https://ui.adsabs.harvard.edu/abs/2007A&A...473..711S} {473, 711}

\bibitem[\protect\citeauthoryear{Sinha \& Garrison}{Sinha \&
  Garrison}{2019}]{Sinha19}
Sinha M.,  Garrison L.,  2019, in Majumdar A.,  Arora R.,  eds, Software
  Challenges to Exascale Computing. Springer Singapore, Singapore, pp 3--20,
  \url {https://doi.org/10.1007/978-981-13-7729-7_1}

\bibitem[\protect\citeauthoryear{{Sinha} \& {Garrison}}{{Sinha} \&
  {Garrison}}{2020}]{Sinha20}
{Sinha} M.,  {Garrison} L.~H.,  2020, \mn@doi [\mnras] {10.1093/mnras/stz3157},
  \href {https://ui.adsabs.harvard.edu/abs/2020MNRAS.491.3022S} {491, 3022}

\bibitem[\protect\citeauthoryear{{Stiskalek}, {Desmond}, {Holvey}  \&
  {Jones}}{{Stiskalek} et~al.}{2021}]{Stiskalek21}
{Stiskalek} R.,  {Desmond} H.,  {Holvey} T.,   {Jones} M.~G.,  2021, \mn@doi
  [\mnras] {10.1093/mnras/stab1845}, \href
  {https://ui.adsabs.harvard.edu/abs/2021MNRAS.506.3205S} {506, 3205}

\bibitem[\protect\citeauthoryear{{Sugiyama} et~al.,}{{Sugiyama}
  et~al.}{2023}]{Sugiyama23}
{Sugiyama} S.,  et~al., 2023, \mn@doi [arXiv e-prints]
  {10.48550/arXiv.2304.00705}, \href
  {https://ui.adsabs.harvard.edu/abs/2023arXiv230400705S} {p. arXiv:2304.00705}

\bibitem[\protect\citeauthoryear{{Takahashi}, {Sato}, {Nishimichi}, {Taruya}
  \& {Oguri}}{{Takahashi} et~al.}{2012}]{Takahashi12}
{Takahashi} R.,  {Sato} M.,  {Nishimichi} T.,  {Taruya} A.,   {Oguri} M.,
  2012, \mn@doi [\apj] {10.1088/0004-637X/761/2/152}, \href
  {http://adsabs.harvard.edu/abs/2012ApJ...761..152T} {761, 152}

\bibitem[\protect\citeauthoryear{{Tonnesen} \& {Ostriker}}{{Tonnesen} \&
  {Ostriker}}{2021}]{Tonnesen21}
{Tonnesen} S.,  {Ostriker} J.~P.,  2021, \mn@doi [\apj]
  {10.3847/1538-4357/ac0724}, \href
  {https://ui.adsabs.harvard.edu/abs/2021ApJ...917...66T} {917, 66}

\bibitem[\protect\citeauthoryear{{Wechsler} \& {Tinker}}{{Wechsler} \&
  {Tinker}}{2018}]{WechslerTinker18}
{Wechsler} R.~H.,  {Tinker} J.~L.,  2018, \mn@doi [\araa]
  {10.1146/annurev-astro-081817-051756}, \href
  {https://ui.adsabs.harvard.edu/abs/2018ARA&A..56..435W} {56, 435}

\bibitem[\protect\citeauthoryear{{Wechsler}, {Bullock}, {Primack}, {Kravtsov}
  \& {Dekel}}{{Wechsler} et~al.}{2002}]{Wechsler02}
{Wechsler} R.~H.,  {Bullock} J.~S.,  {Primack} J.~R.,  {Kravtsov} A.~V.,
  {Dekel} A.,  2002, \mn@doi [\apj] {10.1086/338765}, \href
  {https://ui.adsabs.harvard.edu/abs/2002ApJ...568...52W} {568, 52}

\bibitem[\protect\citeauthoryear{{Wechsler}, {Zentner}, {Bullock}, {Kravtsov}
  \& {Allgood}}{{Wechsler} et~al.}{2006}]{Wechsler06}
{Wechsler} R.~H.,  {Zentner} A.~R.,  {Bullock} J.~S.,  {Kravtsov} A.~V.,
  {Allgood} B.,  2006, \mn@doi [\apj] {10.1086/507120}, \href
  {https://ui.adsabs.harvard.edu/abs/2006ApJ...652...71W} {652, 71}

\bibitem[\protect\citeauthoryear{{Weinberger} et~al.,}{{Weinberger}
  et~al.}{2017}]{Weinberger17}
{Weinberger} R.,  et~al., 2017, \mn@doi [\mnras] {10.1093/mnras/stw2944}, \href
  {https://ui.adsabs.harvard.edu/abs/2017MNRAS.465.3291W} {465, 3291}

\bibitem[\protect\citeauthoryear{{Yang}, {Mo}, {van den Bosch}, {Zhang}  \&
  {Han}}{{Yang} et~al.}{2012}]{Yang12}
{Yang} X.,  {Mo} H.~J.,  {van den Bosch} F.~C.,  {Zhang} Y.,   {Han} J.,  2012,
  \mn@doi [\apj] {10.1088/0004-637X/752/1/41}, \href
  {https://ui.adsabs.harvard.edu/abs/2012ApJ...752...41Y} {752, 41}

\bibitem[\protect\citeauthoryear{{Yu} et~al.,}{{Yu} et~al.}{2022}]{Yu22}
{Yu} J.,  et~al., 2022, \mn@doi [\mnras] {10.1093/mnras/stac2176}, \href
  {https://ui.adsabs.harvard.edu/abs/2022MNRAS.516...57Y} {516, 57}

\bibitem[\protect\citeauthoryear{{Yuan}, {Eisenstein}  \& {Leauthaud}}{{Yuan}
  et~al.}{2020}]{Yuan20}
{Yuan} S.,  {Eisenstein} D.~J.,   {Leauthaud} A.,  2020, \mn@doi [\mnras]
  {10.1093/mnras/staa634}, \href
  {https://ui.adsabs.harvard.edu/abs/2020MNRAS.493.5551Y} {493, 5551}

\bibitem[\protect\citeauthoryear{{van den Bosch}, {Aquino}, {Yang}, {Mo},
  {Pasquali}, {McIntosh}, {Weinmann}  \& {Kang}}{{van den Bosch}
  et~al.}{2008}]{vdB08}
{van den Bosch} F.~C.,  {Aquino} D.,  {Yang} X.,  {Mo} H.~J.,  {Pasquali} A.,
  {McIntosh} D.~H.,  {Weinmann} S.~M.,   {Kang} X.,  2008, \mn@doi [\mnras]
  {10.1111/j.1365-2966.2008.13230.x}, \href
  {https://ui.adsabs.harvard.edu/abs/2008MNRAS.387...79V} {387, 79}

\bibitem[\protect\citeauthoryear{{van den Bosch}, {Jiang}, {Hearin},
  {Campbell}, {Watson}  \& {Padmanabhan}}{{van den Bosch} et~al.}{2014}]{vdB14}
{van den Bosch} F.~C.,  {Jiang} F.,  {Hearin} A.,  {Campbell} D.,  {Watson} D.,
    {Padmanabhan} N.,  2014, \mn@doi [\mnras] {10.1093/mnras/stu1872}, \href
  {https://ui.adsabs.harvard.edu/abs/2014MNRAS.445.1713V} {445, 1713}

\bibitem[\protect\citeauthoryear{{van den Bosch}, {Ogiya}, {Hahn}  \&
  {Burkert}}{{van den Bosch} et~al.}{2018}]{vdB18}
{van den Bosch} F.~C.,  {Ogiya} G.,  {Hahn} O.,   {Burkert} A.,  2018, \mn@doi
  [\mnras] {10.1093/mnras/stx2956}, \href
  {https://ui.adsabs.harvard.edu/abs/2018MNRAS.474.3043V} {474, 3043}

\makeatother
\end{thebibliography}





\appendix

\section{Comparisons between the $\Mprog$ and $\vpeak$ models}
\label{sec:appendix}
\begin{table}
	\centering
	\caption{Summary of the $\zprog$ values in the $\Mprog$ model which gives the best matched-2PCFs to the $\vpeak$ model, and $\langle z_{\rm peak}\rangle$ for the $\vpeak$ model at $z=0,~0.5$ and $1$ for the three samples.}
	\label{tab:best_zprog}
	\begin{tabular}{ccccc} 
		\hline
		~& $n_{\rm gal}~[h^3~{\rm Mpc}^{-3}]$ & $z=0$ & $z=0.5$ & $z=1$\\
		\hline
		~& $10^{-2}$ & $1.12$ & $1.43$ & $1.77$\\
		$\zprog$ & $10^{-3}$ & $0.63$ & $1.03$ & $1.54$\\
		~& $10^{-4}$ & $0.36$ & $0.86$ & $1.43$\\
		\hline
		~& $10^{-2}$ & $1.07$ & $1.35$ & $1.75$\\
		$\langle z_{\rm peak}\rangle$ & $10^{-3}$ & $0.68$ & $1.04$ & $1.50$\\
		~& $10^{-4}$ & $0.45$ & $0.86$ & $1.35$\\
		\hline
	\end{tabular}
\end{table}
\begin{table}
	\centering
	\caption{Summary of the matching rates between the subhalo samples from the $\Mprog$ model with the best match $\zprog$ for each sample and from the $\vpeak$ model at $z=0,~0.5$ and $1$.}
	\label{tab:match_rate}
	\begin{tabular}{cccc} 
		\hline
		$n_{\rm gal}~[h^3~{\rm Mpc}^{-3}]$ & $z=0$ & $z=0.5$ & $z=1$\\
		\hline
		$10^{-2}$ & $94.6\%$ & $94.3\%$ & $94.1\%$\\
		$10^{-3}$ & $94.3\%$ & $93.8\%$ & $93.2\%$\\
		$10^{-4}$ & $93.4\%$ & $92.5\%$ & $91.7\%$\\
		\hline
	\end{tabular}
\end{table}
\begin{figure}
    \includegraphics[width=\columnwidth]{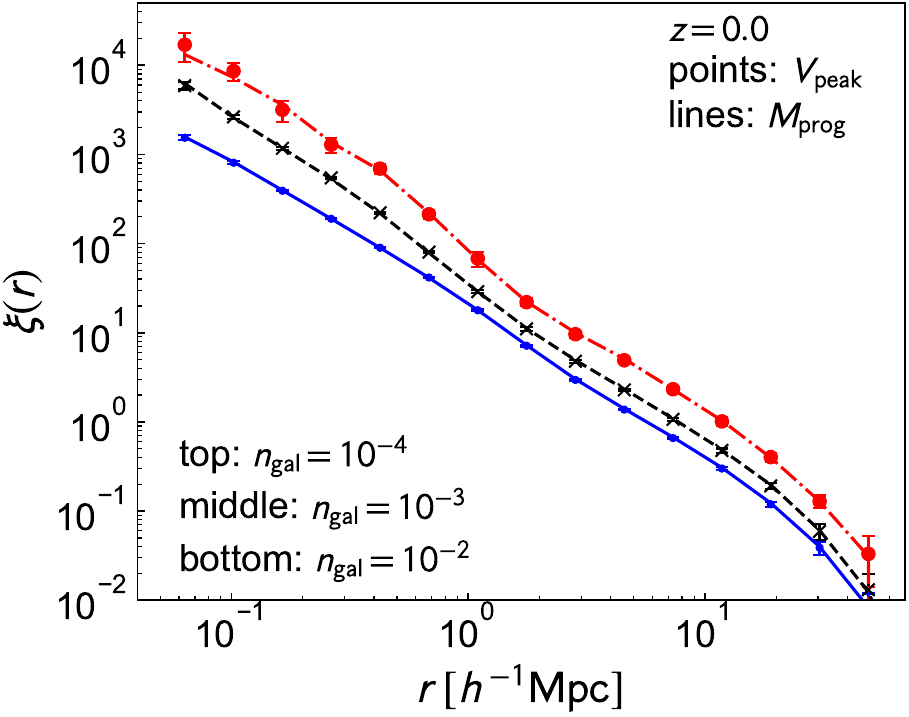}
    \includegraphics[width=\columnwidth]{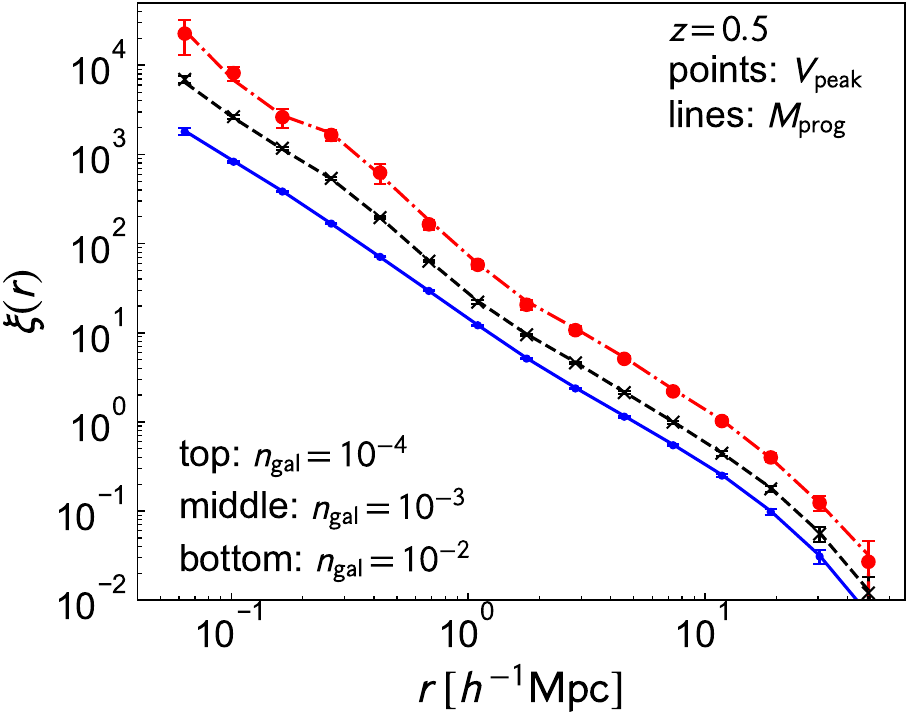}
    \includegraphics[width=\columnwidth]{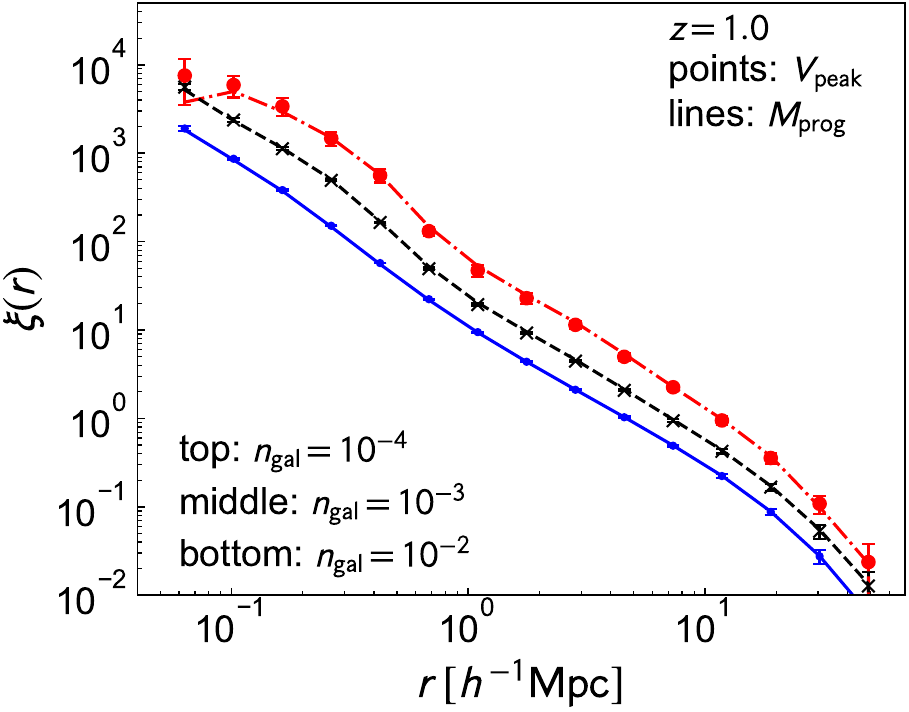}
    \caption{Comparisons between $\xi_M$ and $\xi_V$ for the three samples at $z=0,~0.5$ and $1$.
    For the $\Mprog$ model, the best match $\zprog$ is adopted for each sample.}
    \label{fig:xi_Vpeak_Mprog}
\end{figure}

We here compare the 2PCFs given by the $\Mprog$ and $\vpeak$ models denoted as $\xi_M$ and $\xi_V$, respectively.
As in Sec. \ref{sec:xi_zprog}, we measure $\xi_M$ and $\xi_V$ for the subhalo samples with the number densities of $n_{\rm gal}=10^{-2},~10^{-3}$ and $10^{-4}~h^3~{\rm Mpc}^{-3}$ at $z=0,~0.5$ and $1$.
We seek $\zprog$ that gives the best matched-$\xi_M$ to $\xi_V$. 
For simplicity, we do not perturb either $\Mprog$ or $\vpeak$.

We find that  $\xi_M$ with a certain $\zprog$ value matches with $\xi_V$ very well for all three samples at all three redshifts.
Table \ref{tab:best_zprog} summarizes the best match $\zprog$ values for each sample at $z=0,~0.5$ and $1$.
Fig. \ref{fig:xi_Vpeak_Mprog} compares $\xi_M$ with the best match $\zprog$ and $\xi_V$ for the three samples at $z=0,~0.5$ and $1$, and shows a fairly good agreement between them.

This nice agreement is understood by the mean redshift at which $\vpeak$ is achieved, $\langle z_{\rm peak} \rangle$.
Table \ref{tab:best_zprog} also summarizes the $\langle z_{\rm peak} \rangle$ values for the samples at the redshifts.
We find that the best match $\zprog$ and $\langle z_{\rm peak}\rangle$ are very close to each other within a difference of less than $z=0.1$.
Hence the $\vpeak$ model is equivalent to selecting most massive subhalos at a somewhat higher redshift.
We also compute the matching rate between the subhalo samples from the $\Mprog$ model with the best match $\zprog$ and ones from the $\vpeak$ model.
The matching rates are summarized in Table \ref{tab:match_rate}.
The $\Mprog$ model can select more than $90\%$ of subhalos which are selected by the $\vpeak$ model by taking a certain $\zprog$ value.
Thus the $\Mprog$ model can mimic the $\vpeak$ model in predicting galaxy clustering.

\bsp	
\label{lastpage}
\end{document}